\documentclass{article}

\usepackage{microtype}
\usepackage{graphicx}
\usepackage{booktabs} 

\usepackage[hidelinks]{hyperref}
\usepackage{subfig}

\usepackage[preprint]{icml}

\usepackage{amsmath}
\usepackage{amssymb}
\usepackage{mathtools}
\usepackage{amsthm}

\usepackage{wrapfig}

\usepackage[capitalize,noabbrev]{cleveref}

\theoremstyle{plain}

\theoremstyle{definition}

\theoremstyle{remark}

\usepackage[textsize=tiny]{todonotes}

\icmltitlerunning{MEDIATE: Mutually Endorsed Distributed Incentive Acknowledgment Token Exchange}

\begin{document}

\twocolumn[
\icmltitle{MEDIATE: Mutually Endorsed Distributed Incentive \\ Acknowledgment Token Exchange}




\begin{icmlauthorlist}
\icmlauthor{Philipp Altmann}{1}
\icmlauthor{Katharina Winter}{1}
\icmlauthor{Michael Kölle}{1}
\icmlauthor{Maximilian Zorn}{1}
\icmlauthor{Thomy Phan}{2}
\icmlauthor{Claudia Linnhoff-Popien}{1}
\end{icmlauthorlist}

\icmlaffiliation{1}{LMU Munich, Germany}
\icmlaffiliation{2}{University of Southern California, Los Angeles, USA}

\icmlcorrespondingauthor{Philipp Altmann}{philipp.altmann@ifi.lmu.de}

\icmlkeywords{Multi-Agent Systems, Reinforcement Learning, Peer Incentivization, Consensus, Emergent Cooperation}

\vskip 0.3in
\vskip 0.3in 

] 

\printAffiliationsAndNotice{} 

\begin{abstract} 
Recent advances in \textit{multi-agent systems} (MAS) have shown that incorporating \textit{peer incentivization} (PI) mechanisms vastly improves cooperation. 
Especially in social dilemmas, communication between the agents helps to overcome sub-optimal Nash equilibria. 
However, incentivization tokens need to be carefully selected. 
Furthermore, real-world applications might yield increased privacy requirements and limited exchange. 
Therefore, we extend the PI protocol for \textit{mutual acknowledgment token exchange} (MATE) and provide additional analysis on the impact of the chosen tokens. 
Building upon those insights, we propose \textit{mutually endorsed distributed incentive acknowledgment token exchange} (MEDIATE), an extended PI architecture employing automatic token derivation via decentralized consensus. 
Empirical results show the stable agreement on appropriate tokens yielding superior performance compared to static tokens and state-of-the-art approaches in different social dilemma environments with various reward distributions.
\end{abstract}

\section{Introduction}

Recent advances in using reinforcement learning (RL) in multi-agent systems (MAS) demonstrated their feasibility for real-world multi-agent reinforcement learning (MARL) applications.
Those applications range from smart grids \cite{omitaomu2021artificial} and factories \cite{kim2020multi} to intelligent transportation systems \cite{qureshi2013survey}.
To assess the agents' cooperation capabilities, social dilemmas producing tensions between the individual and collective reward maximization (social welfare) are often used \cite{dawes1980social}. 
Yet, the availability of communication and exchange is vital to fostering cooperation between self-interested individuals.
However, besides the autonomous interaction with an environment, increased privacy requirements might require instances to conceal information regarding their current state \cite{tawalbeh2020iot}.  
Peer incentivization (PI) is a recent branch of research offering a distinct solution for emergent cooperation between agents. 
At its core, PI enables agents to shape each other's behavior by exchanging reward tokens in addition to the environmental reward \cite{phan2022emergent, lupu2020gifting}. 
However, for proper integration and effective incentivization, those exchanged tokens need to be carefully considered, regardless of whether their value is set dynamic or as a hyperparameter.
For the robust and scalable applicability of PI mechanisms in decentralized learning scenarios, adaptive incentivization tokens and mechanisms to agree upon common token values are required. 
Yet, current approaches are missing said coordinated adaptability. 
To overcome these shortcomings, we provide the following contributions:
\begin{itemize}
\item We evaluate the effect of different centralized (common) and decentralized (varying) values for the incentivization token.
\item We propose \textit{mutually endorsed distributed incentive acknowledgment token exchange} (MEDIATE), a mechanism for automatic token derivation based on the agents’ value estimate, and a consensus mechanism to mediate a global token maintaining local privacy. 
\item We provide ablation studies of the introduced token derivation and the consensus mechanism over using a static token. 
Benchmark comparisons to state-of-the-art PI approaches show that MEDIATE can negotiate appropriate tokens that yield improved cooperation and social welfare in various social dilemmas with different reward landscapes. 
\end{itemize}

\section{Background}

\subsection{Problem Formulation}

We formulate our problem of a MAS as a \textit{stochastic game} $\mathcal{M} = \langle \mathcal{D}, \mathcal{S}, \mathcal{Z}, \mathcal{A}, \mathcal{P}, \mathcal{R} \rangle$, with the set of all agents $\mathcal{D} = \{ 1,\dots,N\}$, a set $\mathcal{S}$ of states $s_t$ at time step $t$, a set $\mathcal{A} = \langle\mathcal{A}_{1},\dots, \mathcal{A}_{N}\rangle$ of joint actions $a_t = \langle a_{t,i}\rangle_{i\in\mathcal{D}}$, the transition probability $\mathcal{P}(s_{t+1}\mid s_t, a_t)$, and the joint reward $\mathcal{R}(s_t,a_t)=\langle r_{t,i} \rangle_{i\in\mathcal{D}}\in \mathbb{R}$.
Furthermore, we assume each agent $i$ to have a neighborhood $\mathcal{N}_{t,i} \subseteq \mathcal{D} \setminus \{ i \}$, bounding its set of local observations $z_{t+1}=\langle z_{t+1,i}\rangle_{i\in\mathcal{D}}\in\mathcal{Z}^N$, and the agents' experience tuple $\langle \tau_{t,i}, a_{t,i}, r_{t,i}, z_{t+1,i} \rangle$, where $\tau_{t,i} \in (\mathcal{Z} \times \mathcal{A}_{i})_t$ is the agent's history. 
Agent $i$ selects the next action based on a stochastic policy $\pi_{i}(a_{t,i}|\tau_{t,i})$.
Simultaneously learning agents cause non-stationary, i.e., varying transition probabilities over time.
The goal of each self-interested agent $i$ is to find a \textit{best response} $\pi^{*}_i$ that maximizes the expected individual discounted return: \begin{align}\label{eq:return}  G_{t,i} = \sum_{k=0}^\infty \gamma^k r_{i,t+k},\end{align} with a discount factor $\gamma \in [0,1)$. 
From the perspective of an agent, other agents are part of its environment, and policy updates by other agents affect the performance of an agent's own policy \cite{laurent2011world}. 
The performance of $\pi_i$ is evaluated using a \textit{value function} $V_i(s_{t}) = \mathbb{E}_{\pi}[ G_{t,i} | s_t ]$ for all $s_t\in\mathcal{S}$, with the \textit{joint policy} $\pi =\langle\pi_j\rangle_{j\in\mathcal{D}}$ \cite{bucsoniu2010multi}. 
Both the policies $\pi$ and the value functions $V$ are approximated by independent neural networks parameterized by $\theta$ and $\omega$ respectively. 
For simplicity we omit those for the following and use the abbreviated forms $V_i = V_i^\omega(\tau_{t,i}) \approx V^{\pi_i}(s_t)$ and $\pi_i = \pi^\theta_i$ respectively.
To measure \textit{efficiency} $U$ of the whole MAS we furthermore consider the social welfare \cite{sandholm1996multiagent}, measured by the sum of undiscounted returns over all agents within an episode until time step $T$: 
\begin{align}\label{eq:efficiency} U = \sum_{i\in\mathcal{D}} \sum_{t=0}^{T-1} r_{t,i} \end{align}

\subsection{Social Dilemmas}

Game Theory analyzes behavior among rational agents in cooperative and competitive situations \cite{russell2010artificial, littman2001value}.
Social dilemmas are Markov games that inhibit a specific reward structure, which creates tension between individual and collective reward maximization.
\textit{Sequential social dilemmas} (SSD) are temporally extended social dilemmas, in which the game repeats over several time steps \cite{leibo2017multi}.
The Nash equilibrium is a situation where no agent can increase its individual reward by changing its strategy if all other agents maintain their current strategy \cite{littman2001value, sandholm1996multiagent}.
MARL utilizes SSDs to analyze and experiment with the social behavior of different learning strategies \cite{leibo2017multi}. 
To assess the emergence of cooperation, we employ the \textit{Iterated Prisoner's Dilemma} (IPD), where mutual defection constitutes a Nash equilibrium \cite{axelrod1980effective,sandholm1996multiagent}. 
To evaluate the scalability of our approach, we use the \textit{Coin Game} with two, four, and six agents \cite{lerer2017maintaining}. 
Additionally, we use the \textit{Rescaled Coin Game} with two agents to assess the robustness w.r.t. varying reward landscapes.
The rate of \textit{own coins} versus total coins collected reflects overall cooperation.
For insights on long-term cooperation, we use \textit{Harvest}, posing a risk of the \textit{tragedy of the commons} to self-interested agents \cite{perolat2017multiagent,phan2022emergent}. 
For further details about the employed environments, please refer to \autoref{sec:envs}.

\subsection{Peer Incentivization}
In MAS, cooperation connotes the joining of individual problem-solving strategies of autonomous agents into a combined strategy \cite{crainic2007explicit}. 
The emergent cooperation of learning agents necessitates coordination \cite{noe2006cooperation}, which poses a vital challenge to current communication protocols in decentralized MARL scenarios \cite{jaques2019social}. 
PI is a recent branch of research, focussing on agents learning to actively shape the behavior of others by sending rewards or penalties \cite{phan2022emergent, yang2020learning}. 
These peer rewards are processed like environment rewards, enabling the emergence of cooperation. 
However, new dynamics arise through the increased inter-dependency, which comes with new challenges. 
Carefully designing this reward mechanism is essential to achieving a good outcome \cite{lupu2020gifting}.

\subsection{Consensus in Multi-Agent Systems}\label{subsec:Consensus}
Distributed systems use consensus algorithms to deduct a global average of local information \cite{schenato2007distributed}. 
For MAS, consensus describes the convergence of agents on a mutual value via communication \cite{Li2019ASO}.
A consensus algorithm specifies the execution steps to reach consensus \cite{han2013cluster}.
Bee swarms, bird flocks, and other group-coordinated species show natural behavior \cite{amirkhani2022consensus} that inspires further underlying concepts like leadership, voting, or decision-making \cite{conradt2005consensus}.
Two main application areas for consensus algorithms are sensor networks \cite{yu2009distributed} and blockchain technology \cite{monrat2019survey}, which has played an integral role in cryptocurrencies and provides promising solutions for IoT applications. 
Consensus in sensor networks mainly deals with the fusion of distributed data, especially for time-critical data \cite{schenato2007distributed} and uncertainty in large-scale networks \cite{olfati2005consensus}.
Research in cryptocurrency and IoT focuses on synchronization \cite{cao2019internet}, agreement \cite{salimitari2018survey}, and verification of actions \cite{lashkari2021comprehensive} between entities in distributed systems.  
The number of sophisticated consensus algorithms is growing through the rising importance of decentralized coordination mechanisms \cite{lashkari2021comprehensive} in an increasingly digitally connected world.
Our approach utilizes the cryptographic technique of additive secret sharing, solving the average consensus problem for privacy-critical tasks \cite{li2019privacy}. 
MARL research on consensus algorithms has been increasing recently, intending to reach an optimal joint policy in a decentralized system that is robust to unreliable agents or adversarial attacks \cite{figura2021adversarial}.
To our knowledge, no research exists concerning consensus algorithms, PI and RL.

\section{Related Work}
Various concepts affect the achievement of emergent cooperation in MAS. 
\textit{Learning with opponent-learning awareness} (LOLA) \cite{Foerster2017Learning} and \textit{stable opponent shaping} (SOS) \cite{letcher2018stable} consider the learning process of other agents and shape the policy updates of opponents. 
Nature and human social behavior also inspired many concepts. 
\citet{Wang2018Evolving} developed an evolutionary approach to create agents with social behavior by natural selection. 
Other work focuses on prosocial agents and intrinsic motivation thriving for the manifestation of social norms \cite{jaques2019social}. 
\citet{eccles2019learning} divided agents into innovators, learning a policy, and imitators, which reciprocate innovators. 
\citet{baumann2020adaptive} insert an external planning agent into the environment, which can observe all agents and distribute rewards. 
Overall, we divide approaches fostering emergent cooperation into constructed artificial social assemblies, added intrinsic motivation, and external optimization techniques. 
Our approach combines those concepts, using socially inspired mutual acknowledgment to shape the environmental rewards. 

A large corpus in PI research focuses on similar approaches to learning incentives integrated into the model. 
\textit{Gifting} integrates the reward-gifting capability into agents' policies as an additional action. 
Different reward mechanisms can build upon this concept. 
In \textit{zero-sum gifting}, agents receive a penalty for each sent reward to balance the total sum of rewards. 
Gifting can also be only allowed up to a \textit{fixed budget} per episode as an alternative to penalization. 
With a \textit{replenishable budget}, the reception of environment rewards can recharge this budget \cite{lupu2020gifting}. 
\textit{Learning to incentivize other learning agents} (LIO) is another approach that uses an incentive function to learn appropriate peer rewards. 
Selecting a reward is not part of the action space but is learned separately by a second model \cite{yang2020learning}. 
Like LIO, MEDIATE derives incentives from the agents' expected environmental return. 
However, in contrast to MEDIATE, LIO requires an additional model to be learned for predicting this value, causing additional overhead. 
\textit{Learning to share} (LToS) also implements two policies, one for local objectives set by a high-level policy \cite{yi2021learning}. 
\textit{Peer-evaluation-based dual-DQN} (PED-DQN) lets agents evaluate their received peer signals w.r.t. their environment rewards with an additional DQN network \cite{hostallero2020inducing}. 
\textit{Learning to influence through evaluative feedback} (LIEF) learns to reconstruct the reward function of peers via feedback. 
The authors call for an investigation between a manual, systematic, and learned construction of rewards \cite{merhej2021lief}. 
\citet{fayad2021influence} use counterfactual simulations to derive influential actions.  
The above concepts modify the agent models or the action space to derive the intrinsic rewards.
Rather than altering the agents themselves, we utilize an additional protocol layer, serving as a tool for agents, yielding increased flexibility. 

\textbf{MATE}\quad \textit{Mutual acknowledgment token exchange} (MATE) proposed by \citet{phan2022emergent} controls the exchange of incentives via a two-phase communication protocol. 
In the request phase of each time step, all agents evaluate their \textit{monotonic improvement} (MI) and potentially send acknowledgment tokens to all neighbors. 
In the response phase, agents evaluate their MI w.r.t. the sum of environment rewards and the received token and respond with a positive or negative token. 
This two-way handshake allows agents to give feedback to other agents when incentives are received, which fosters cooperation and outperforms naïve learning and other PI approaches, like LIO and Gifting, in various benchmarks regarding efficiency and equality metrics \cite{phan2022emergent}. 
MATE uses a communication layer and thus provides a lightweight solution with minimal interference with the agent model. 
Due to this flexible and privacy-conserving design, we evaluate our approach as an extension of MATE. 
However, note that other protocol PI solutions can also utilize MEDIATE.
Overall, we aim to eliminate the need for setting the exchange token beforehand, which denotes a central limitation of MATE. 

Given their direct combination with the external reward, we argue that incentivization tokens are sensitive parameters to be carefully considered. 
\citet{kuhnle2023learning} analyze the Harsanyi-Shapley value to determine the weight of a side payment based on the strategic strength of a player in two-player scenarios. 
Value decomposition networks \cite{Sunehag2018}, QMIX \cite{rashid2020monotonic}, and QTRAN \cite{son2019qtran} decompose the joint action-value function into agent-based value functions to achieve cooperation and maximize social welfare. 
These approaches are based on a centralized value function, whereas our work focuses on independent learners in a fully decentralized setting.
MEDIATE also uses the value function to automatically derive token values to be mixed with the environmental reward, posing a lightweight and efficient solution.

\begin{figure*}
    \centering \vspace{-.65cm}
    \subfloat[\centering Central token values for CoinGame-2, -4, and -6]{ 
        \includegraphics[width=.5\linewidth]{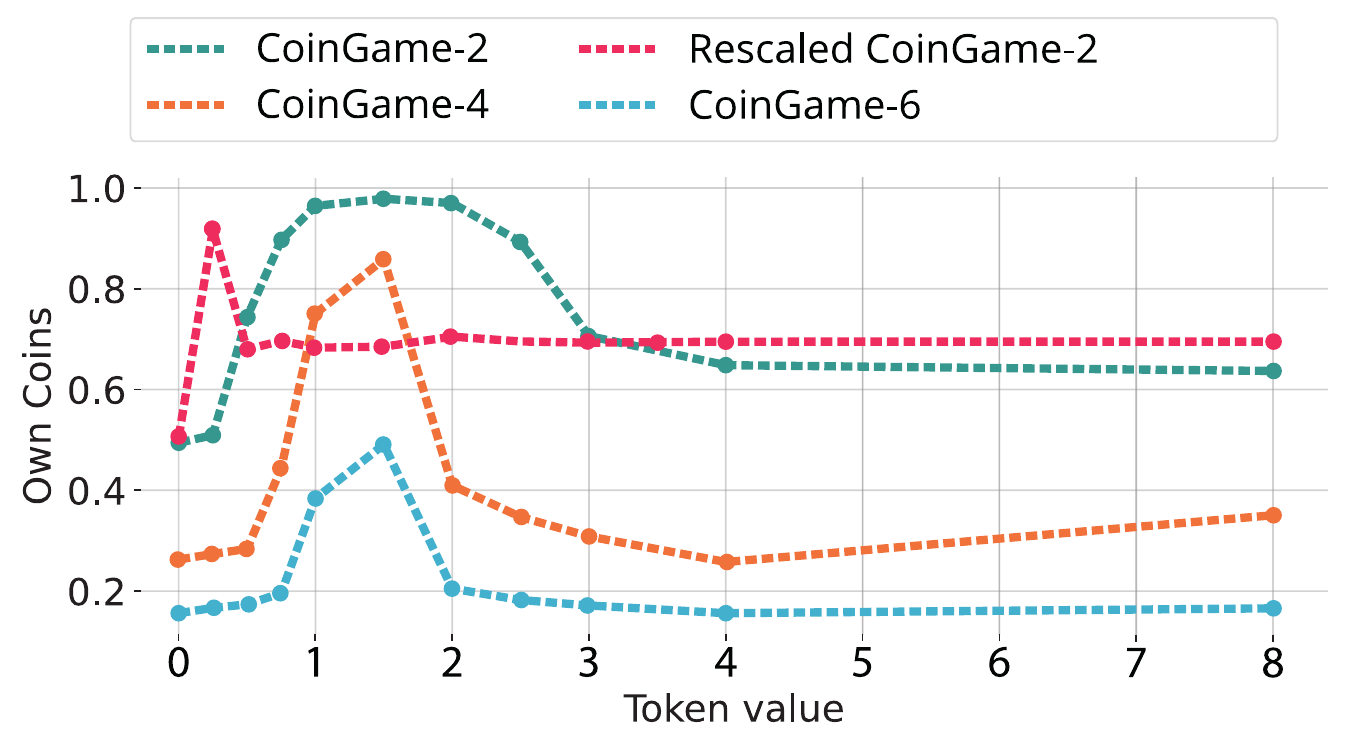}\label{fig:token-analysis:central}}
    \subfloat[\centering Decentralized Token Values for CoinGame-2]{\includegraphics[width=.3\textwidth]{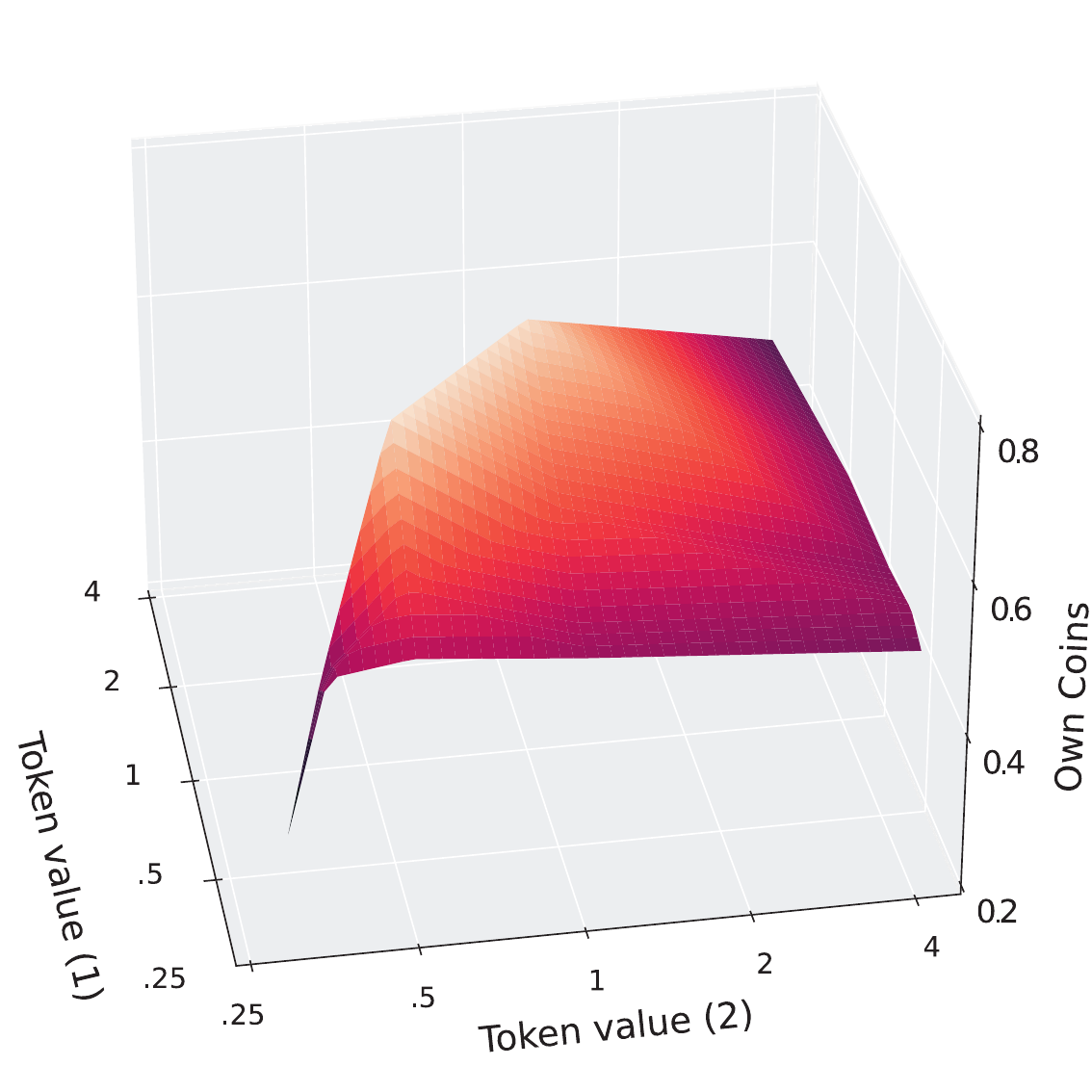}\label{fig:token-analysis:decentral}}
  \caption{Rate of \textit{own coins} for different tokens when determined centralized (\ref{fig:token-analysis:central}) and decentralized (\ref{fig:token-analysis:decentral})}
  \label{fig:token-analysis}
\end{figure*}

\section{Impact of Incentivization Tokens}

As MATE was previously only evaluated with token values of $1$, we first aim to provide additional insights into the impact of the incentivization token, supplying an extensive hyperparameter analysis across various choices, both per-agent (\textit{decentralized}) and globally (\textit{centralized}).
Fig.~\ref{fig:token-analysis:central} displays the level of cooperation measured by the rate of \textit{own coins} collected for different token values $\mathcal{T}\in[0,0.25,0.5,0.75,1,1.5,2,2.5,3,4,8]$ in the Coin Game with two, four, and six agents, as well as the two-agent Coin Game with scaled rewards. 
We averaged all results over five random seeds. 
The graphs display high average levels of cooperation for value 1 in all settings, except for the down-scaled Coin Game, where token 1 fails, indicating that the token value is highly dependent on the reward landscape.
Insufficient (inferior) token values fail to achieve the collective objective, causing self-interested behavior. 
Conversely, over-exploitative (intemperate) token values likewise fail to yield cooperative behavior.
Increasing the number of agents, a value of 1.5 appears to be optimal within the presented range, but the required precision for successful cooperation varies. 
Also the range of token values that yield high cooperation narrows, retaining its relative position but exhibiting an increased sensitivity to the boundaries of that range. 
The discrepancy between the optimal token value of 1.5 and value 1 increases in the six-agent Coin Game.
The analysis implies that factors like the domain, the reward landscape, and the number of agents influence incentive rewards.
The range of tokens with distinctively high cooperation is solely a function of the environment rewards but depends on the specific dynamics of the game, making it challenging to predict.
A fixed token value lacks the adaptability required for diverse settings, making a priori prediction based on parameter settings a complex task. 
It becomes evident that reward structures are not the sole determinants for selecting appropriate token weights and may not even be reliably indicative across all scenarios.

To provide further insights into the dynamics introduced by the choice of incentivization token value, we modified the protocol to allow the agents to exchange disparate tokens.
We refer to this mode as \textit{decentralized}.
Note that using automated token derivation in a decentralized setting without a mechanism for coordination or consensus might result in such varying token values. 
Fig.~\ref{fig:token-analysis:decentral} maps the interpolated cooperation levels in the two-agent Coin Game with the tokens $\mathcal{T}\in[0.25, 0.5, 1, 2, 4]$, as values between 1 and 2 have previously shown to be sufficient central tokens, employed by both agents, measured by the rate of own coins. 
The results reveal that the token combinations (1,1) and (2,2) yield the highest cooperation rates. 
Both token values are positioned in the appropriate token range in the centralized comparison (cf. Fig.~\ref{fig:token-analysis:central}) and the combinations contain equal values, which appears to be a significant criterion in this context. 
Although the combination (1, 2) includes two appropriate values, the cooperation is decreased compared to the equal-valued exchange.
With increasing discrepancy between the token values, cooperation further decreases, suggesting a correlation between the degree of value equality and cooperation. 
Agents with over-exploitative token values can exert high impact on other agents, especially on those with limited social influence due to smaller tokens, leading to a manipulative form of cooperation.  
Equal but inappropriate token values exhibit low performance and cooperation, which minimizes for (0.25, 0.25). 
Overall, this evaluation suggests that the exchange of decentralized token values must be appropriate and equal to provide fairness and induce equal cooperation. 

Nevertheless, the rate of own coins collected for all tested tokens excels the performance of naïve learning, reflected by token value $0$.
Conceptually, these prospects of MATE arise from enabling agents to share their success, provided the benefits are mutual. 
As shown before, however, exchanging tokens of value $\mathcal{T}=1$ might not always be the sufficient choice for any given environment.

\section{MEDIATE}

\begin{figure}[h]
    \centering
    \includegraphics[width=.45\textwidth]{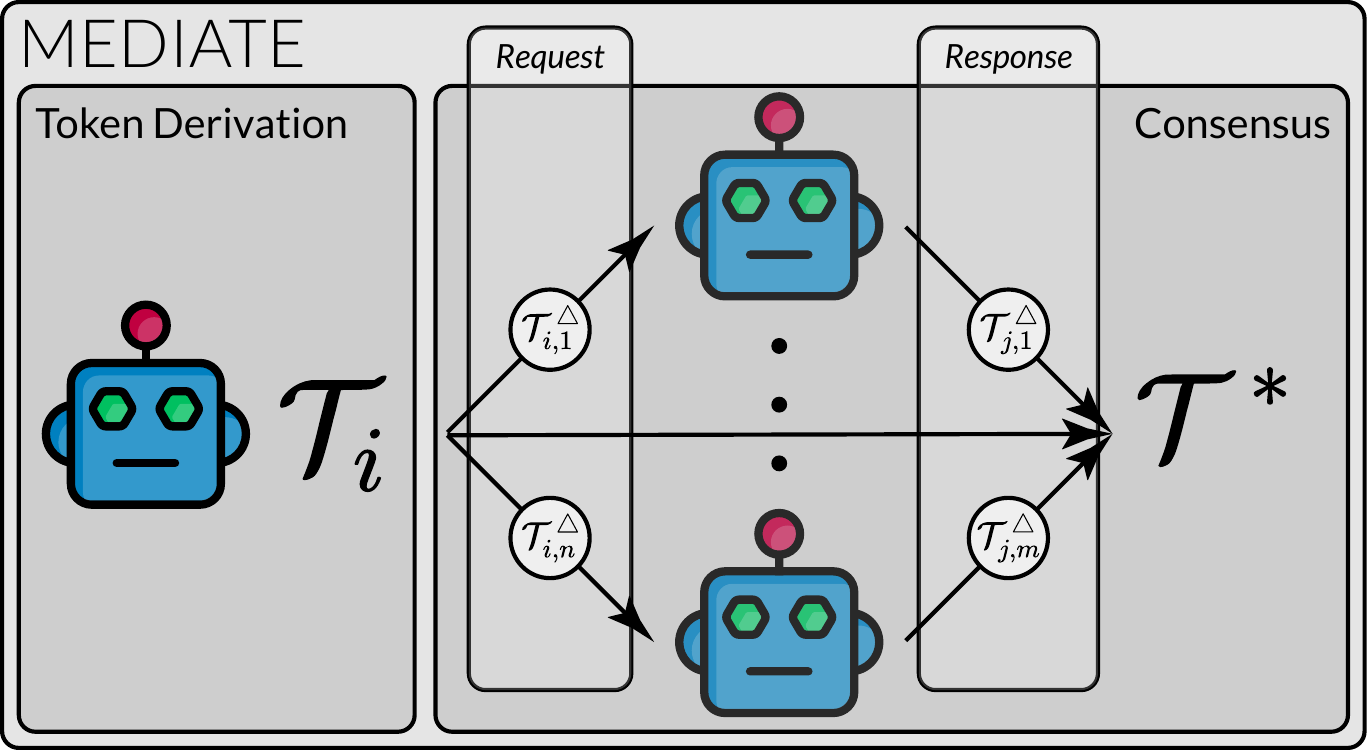}
    \caption{MEDIATE Architecture}
    \label{fig:MEDIATE}
\end{figure}

To elevate PI token values from static hyperparameters to dynamically adaptable domain-specific quantities, we propose \textit{mutually endorsed distributed incentive acknowledgment token exchange} (MEDIATE), combining two progressions (cf. Fig.~\ref{fig:MEDIATE}): 
First, we provide an automated mechanism to derive dynamic agent-based incentivization tokens $\mathcal{T}_i$.
To ensure global convergence of said tokens, we secondly provide a consensus mechanism that ensures the privacy of the agents' local information. 

As the exchanged tokens acknowledge mutual benefit, using $V_i$ to measure the value of a state for agent $i$ is the most natural choice for its automation. 
Furthermore, this allows us to provide a lightweight extension, not relying on additional models to be learned (in contrast to previous automatic incentivization approaches). 
MEDIATE operates decentralized, individually calculating a token value for each agent based on their respective value functions.
An agent does not know the value function of other agents, but we assume overall similar value functions. 
This assumption ensures both independence and decentralization by enabling an agent to operate solely based on its domain-specific metrics and variables.
Alg.~\ref{alg:MEDIATE} depicts the proposed mechanism for deriving and updating individual tokens:
\begin{algorithm}[b]
\renewcommand{\algorithmicrequire}{\textbf{Setup} for Agent $i \in \mathcal{D}$:} 
\caption{Agent-wise Token Derivation with MEDIATE}\label{alg:MEDIATE}
\begin{algorithmic}
\REQUIRE $\mathcal{T}_i \gets 0.1; r_i^{\min}\gets \infty; \widetilde{V}_i \gets 0$ \label{line:token-initialization}
\FOR{Epoch $\epsilon$ in Epochs; Agent \(i \in \mathcal{D} \) }
\STATE $\overline{V}_i\gets \{\}$ \quad $ \triangleright$ \textit{Initialize mean values for epoch}
\FOR{Rollout $\langle \tau_{0,i}, a_{0,i}, r_{0,i}, \dots, \tau_{T,i}, a_{T,i}, r_{T,i} \rangle$ in $\epsilon$} 
\STATE $ r_i^{\min} \gets \min (r_i^{\min}, \langle r_{0\dots T,i}\rangle) $ \label{line:rmin}
\STATE $\overline{V}_i\gets\overline{V}_i \cup  \bar{V}_i(\tau) $ \quad \textit{$\triangleright$ Calculate mean value \eqref{eq:meancumvalue}}\label{line:sum-state-values}
\ENDFOR
\STATE $\mathcal{T}_i \gets \max(\mathcal{T}_{(i)}^{(\ast)} + \nabla_{\mathcal{T}_i},0)$ \quad \textit{$\triangleright$ Update local token \eqref{eq:tokenupdate}}\label{line:token-update} 
\STATE $\widetilde{V}_i \gets \text{median} (\overline{V}_i) $
\ENDFOR
\end{algorithmic}
\end{algorithm}
All agents initially set their token to a small but non-zero value of $0.1$ to differentiate it from a zero-valued token that would equate to naïve learning.
This initialization allows for the immediate incorporation of the MATE mechanism. 
We implemented MEDIATE extending MATE-TD, which employs temporal-difference-based MI evaluation rooted in the value function. 
To ensure an appropriate acceptance-/rejection-ratio and thus an appropriate impact on the behavior of other agents, the token value must be proportional to the value function. 
Thus, we suggest incrementing tokens by the relative difference between the mean state value estimates across consecutive epochs. 
By doing so, MEDIATE tailors tokens to the unique dynamics of each domain, thereby fostering equal cooperation across diverse settings.
As a measure of the profit, we derive the mean accumulated value $\bar{V}$ of an episode $\tau$ of length $T$ similar to the undiscounted return (cf. Eq.~\eqref{eq:return}):
\begin{align}\label{eq:meancumvalue}\bar{V}_i(\tau) = \frac{\sum_{t=0}^T V_i(\tau_{t,i})}{T}\end{align}
$V_i$ refers to the current value approximation of agent $i$.
Furthermore, we use the median of the mean values $\bar{V}$ over an epoch of episodes to improve stability.
The local tokens $\mathcal{T}_i$ are adjusted every epoch based on the difference ($\Delta$) between the current median of the mean values ($\text{median} (\overline{V}_i)$) and the previous median of the mean value $\widetilde{V}_i$: 
\begin{align}\label{eq:tokenupdate}\nabla_{\mathcal{T}_i} = \alpha \cdot \frac{\Delta(\widetilde{V}_i, \text{median} (\overline{V}_i))}{\widetilde{V}_i} \cdot |r_i^{\min}|,\end{align}
with $\alpha=0.1$ as a constant comparable to a learning rate and the absolute value of the lowest encountered environmental reward $r_i^{\min}$ (cf. Alg.~\ref{alg:MEDIATE}) as a scaling factor. 
Furthermore, we use the previous median of the mean value $\widetilde{V}_i$ for normalization.
Consequently, sufficiently large negative state value estimates can cause positive tokens, which rise when the value further decreases. 
For negative values, the token thus remains proportionate to the absolute magnitude of the value function. 
Furthermore, the resulting token value is clamped to positive values using the $\max$ operation (cf. Alg.~\ref{alg:MEDIATE}), sending a zero token otherwise.
Resembling the use of a ReLU activation function \cite{agarap2018deep}, this forces the agent to send no incentive when unable to send a positive. 
By this, agents adhere to the principle of \textit{Niceness}, which is a core principle for the reciprocal strategy of MATE, implying no intent of defection in the request \cite{phan2022emergent}.

However, besides using appropriate tokens, findings from the analysis of decentralized tokens also demonstrated the need for equal token values in the mutual exchange. 
Therefore, we extend MEDIATE with a consensus mechanism to reach an agreement on a mutual token, increasing equality and reducing the impact of outliers while preserving the privacy of the agents' confidential information using additive secret sharing. 
All agents set up the consensus exchange by dividing their token values into shares for all agents in their neighborhood $\mathcal{N}$, reserving one share for privacy reasons. 
\newpage
The token is only reconstructable when accounting for all shares, which provides security against privacy defectors. 
In the request phase, all agents $i$ send the corresponding shares $[\mathcal{T}^\triangle_{i,1}, \dots, \mathcal{T}^\triangle_{i,n}]$ to all $n$ neighbors. 
Each receiving agent $j$ accumulates its received shares $[\mathcal{T}^\triangle_{j,1}, \dots, \mathcal{T}^\triangle_{j,m+1}]$ from its $m$ neighbors, including its reserved share. 
In the response phase, each agent $j$ sends the accumulated shares to all its neighbors. 
Each receiving agent $i$ obtains the accumulated shares from all neighbors, which it averages over the number of shares, i.e., the number of agents $N$, to obtain the reconstructed consensus token $\mathcal{T}^\ast$: 
\begin{align}\mathcal{T}^\ast = \frac{\sum_{i\in\mathcal{N}}{\sum_{j\in\mathcal{N}}{\mathcal{T}^\triangle_{i,j}}}}{N}\end{align}
In domains like Harvest, with only partially connected agents and changing topologies, the consensus protocol includes a multi-iteration response phase. 
Each summed share is tagged with an ID, sent to all neighbors, and forwarded over multiple time steps to ensure network-wide information dissemination. 
To integrate the reconstructed token into the token derivation mechanism, we propose two different update mechanisms:
\textit{Isolated} updates the local token $\mathcal{T}_i$ based on the previous local token, which is shared independently via the consensus protocol: $\max(\mathcal{T}_i + \nabla_{\mathcal{T}_i},0)$. 
In contrast, \textit{synchronized} replaces the local token with the reconstructed token $\mathcal{T}^\ast$ after the consensus phase: $\max(\mathcal{T}^\ast+ \nabla_{\mathcal{T}_i},0)$. 
Consequently, only the token update (cf. Alg.~\ref{alg:MEDIATE}) is affected, either synchronized with the consensus token $\mathcal{T}^\ast$ or drifting independently.
We will refer to the resulting variants as \textit{MEDIATE-I} and \textit{MEDIATE-S}.

\section{Experimental Results}
To assess the effect of the introduced token derivation mechanism and the proposed consensus architecture, we ran evaluations comparing \textit{isolated} and \textit{synchronized} MEDIATE in the \textit{IPD}, \textit{CoinGame-2}, and \textit{CoinGame-4}.
As an additional ablation, we use a reduced version with only the automated decentralized token derivation (cf. Alg.~\ref{alg:MEDIATE}) but without any consensus mechanism, which we refer to as \textit{AutoMATE}. 
Additionally, we compare the above to naïve learning and MATE with a fixed token of $1$. 
We measure cooperation in all Coin Game environments by the ratio between \textit{own coins collected} (\texttt{occ}) and \textit{total coins collected} (\texttt{tcc}):  $\textit{own coins} = \frac{|\texttt{occ}|}{|\texttt{tcc}|}$.
We compare the performance in the \textit{IPD} and \textit{Harvest} by the approaches' \textit{efficiency} (cf. Eq.~\eqref{eq:efficiency}), as a metric for social welfare. 
Additionally, we compare all MEDIATE ablations w.r.t. the convergence of their token value.
To test the scalability of MEDIATE and its robustness to varying reward distributions, we provide further evaluations in the \textit{Rescaled Coin Game-$2$}, \textit{CoinGame-6}, and \textit{Harvest}, including benchmark comparisons to zero-sum- and budget-gifting and LIO. 

Training is conducted for 5000 epochs, comprising ten episodes each.
We averaged all of the following results over eight random seeds.
If not stated otherwise, all implementations use their default hyperparameters from the corresponding source. 
Please refer to the appendix for further environment- and implementation details. 

\subsection{Evaluation of MEDIATE}

\begin{figure*}[t]\centering\includegraphics[width=.8\linewidth]{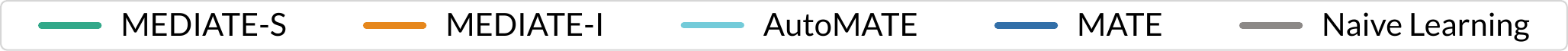}\vspace{-0.35cm}
  \subfloat[\centering IPD | Efficiency]{\includegraphics[width=.33\linewidth]{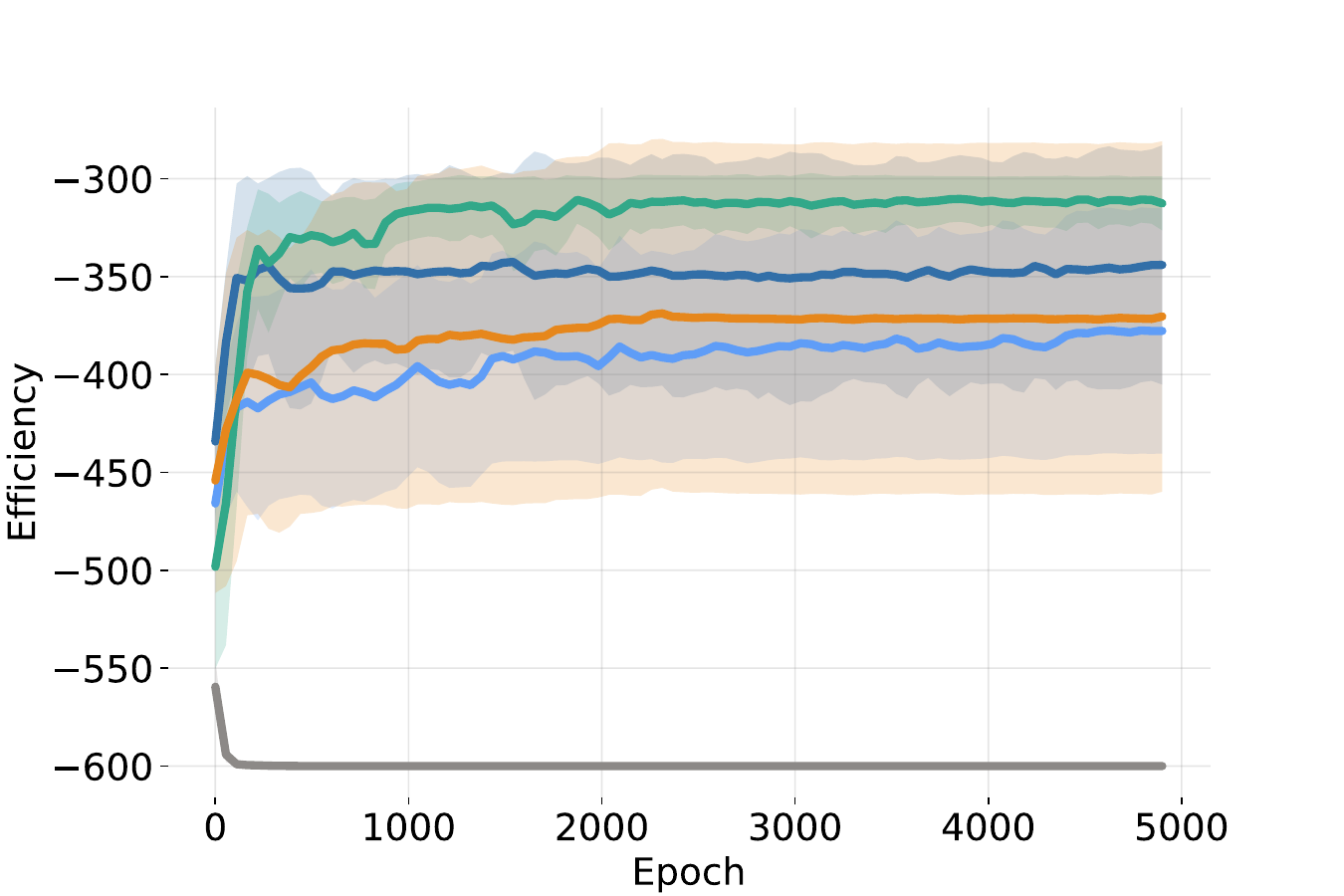}\label{fig:eval:e:ipd}}
  \subfloat[\centering Coin Game (2 agents) | Own Coins]{\includegraphics[width=.33\linewidth]{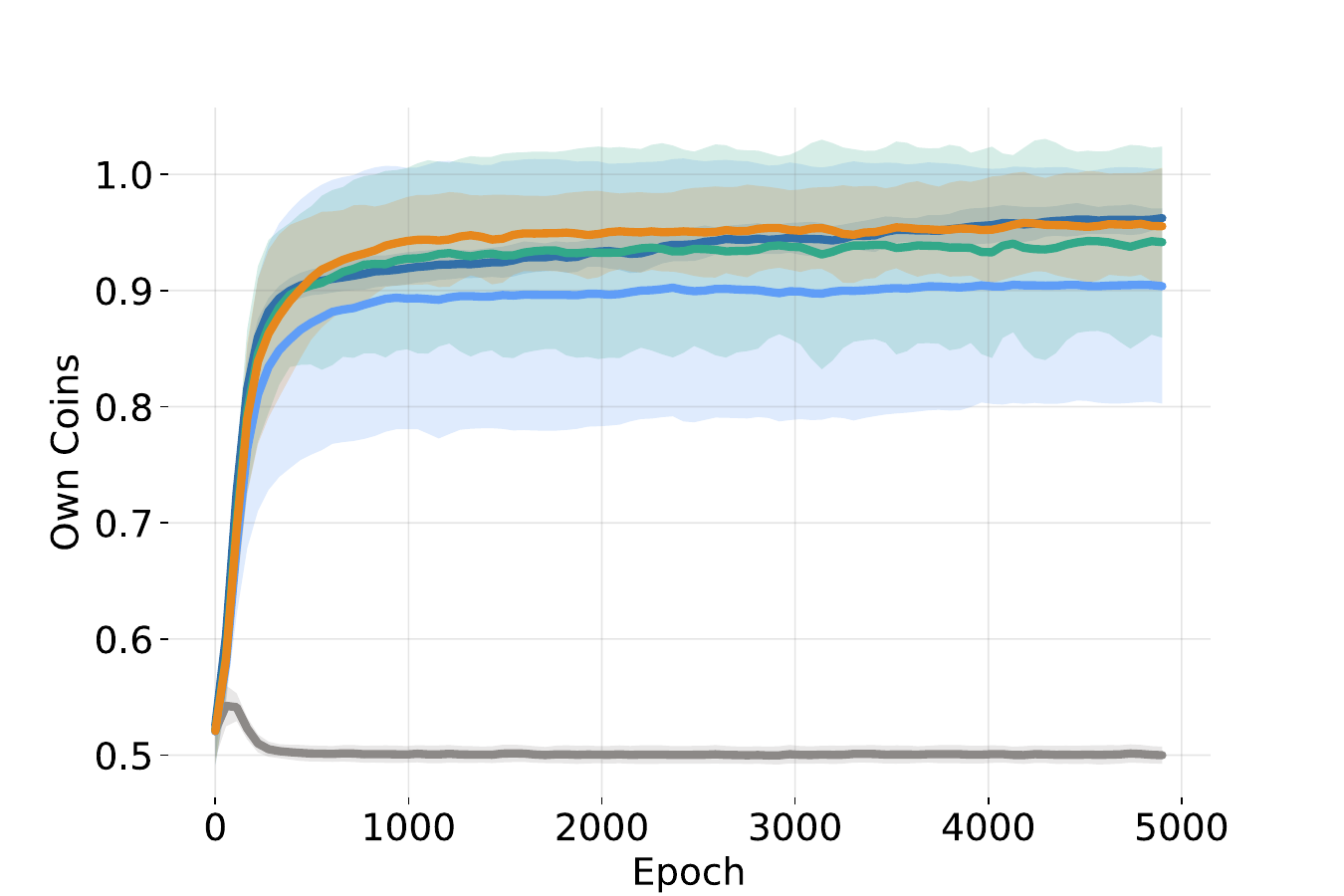}\label{fig:eval:e:coin2}}
  \subfloat[\centering Coin Game (4 agents) | Own Coins]{\includegraphics[width=.33\linewidth]{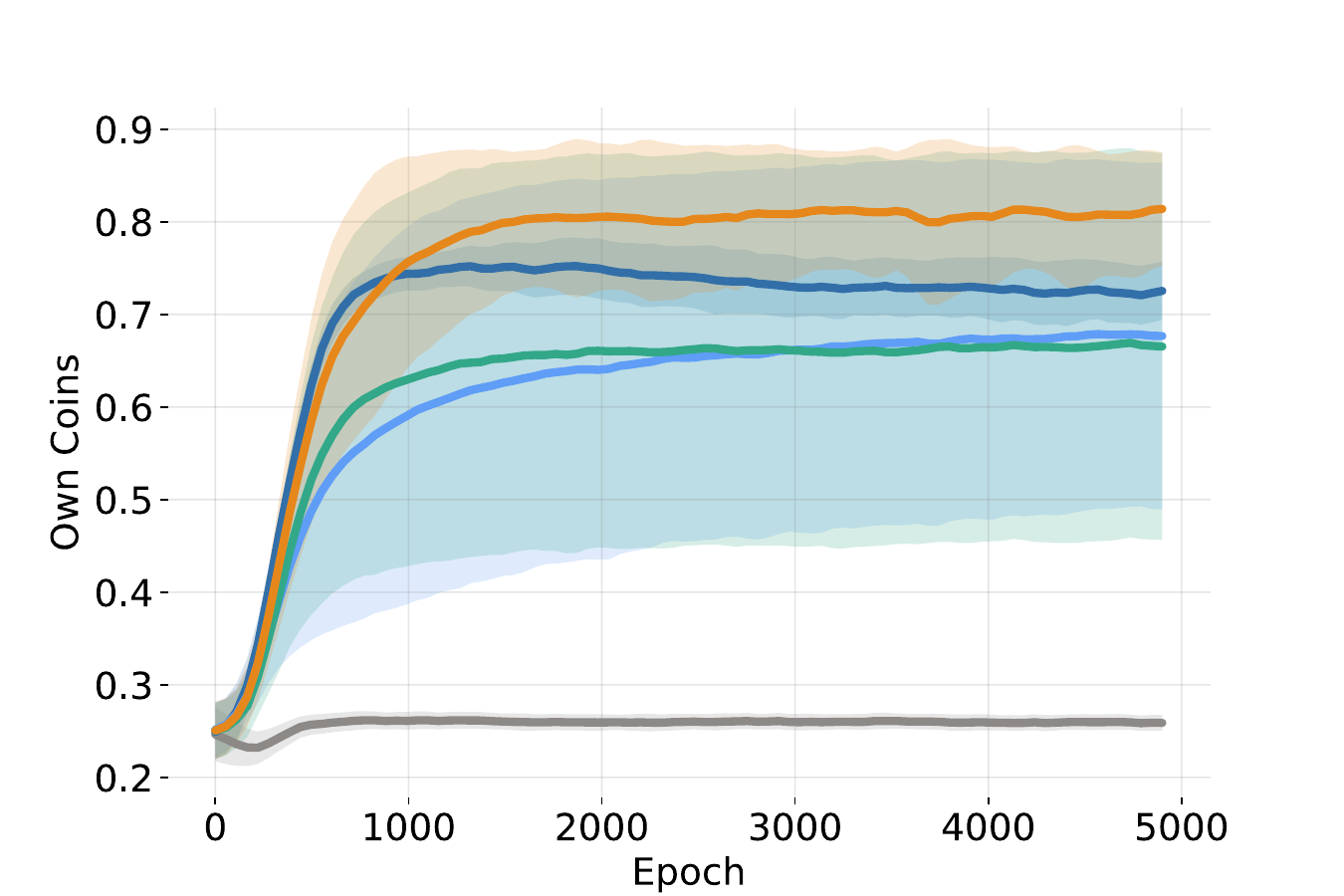}\label{fig:eval:e:coin4}}\\
  \subfloat[\centering IPD | Token Values]{\includegraphics[width=.33\linewidth]{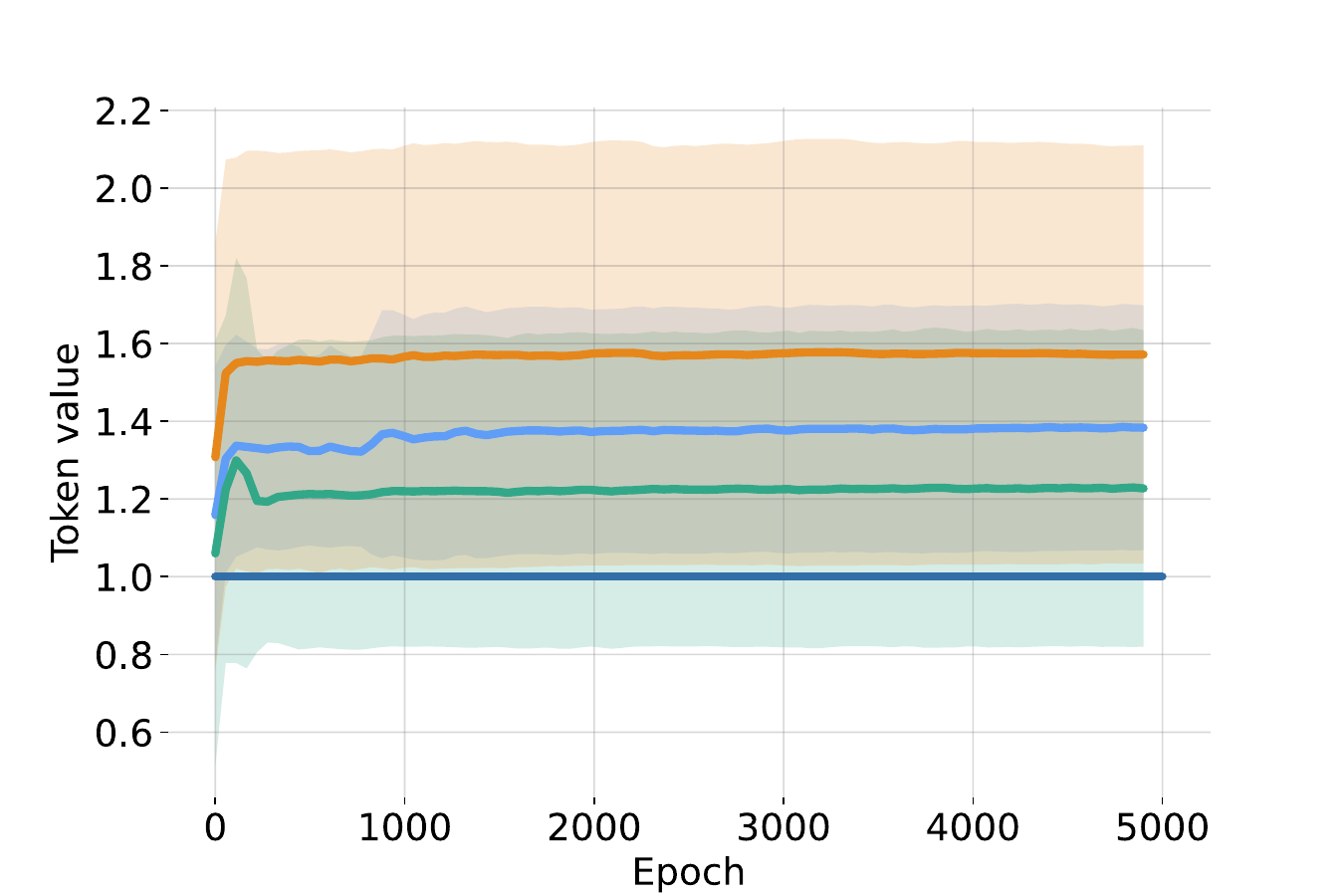}\label{fig:eval:t:ipd}}
  \subfloat[\centering Coin Game (2 agents) | Token Values]{\includegraphics[width=.33\linewidth]{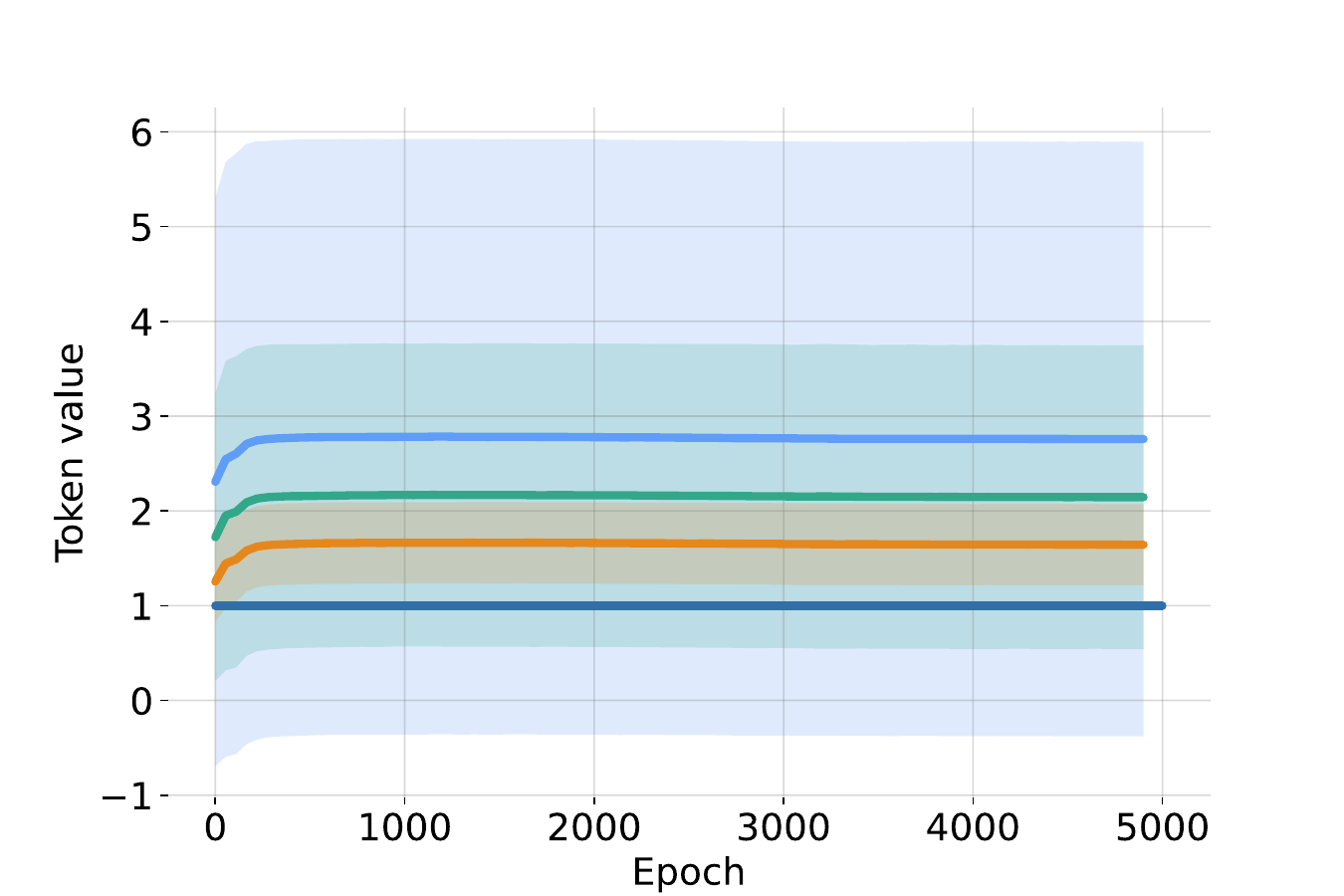}\label{fig:eval:t:coin2}}
  \subfloat[\centering Coin Game (4 agents) | Token Values]{\includegraphics[width=.33\linewidth]{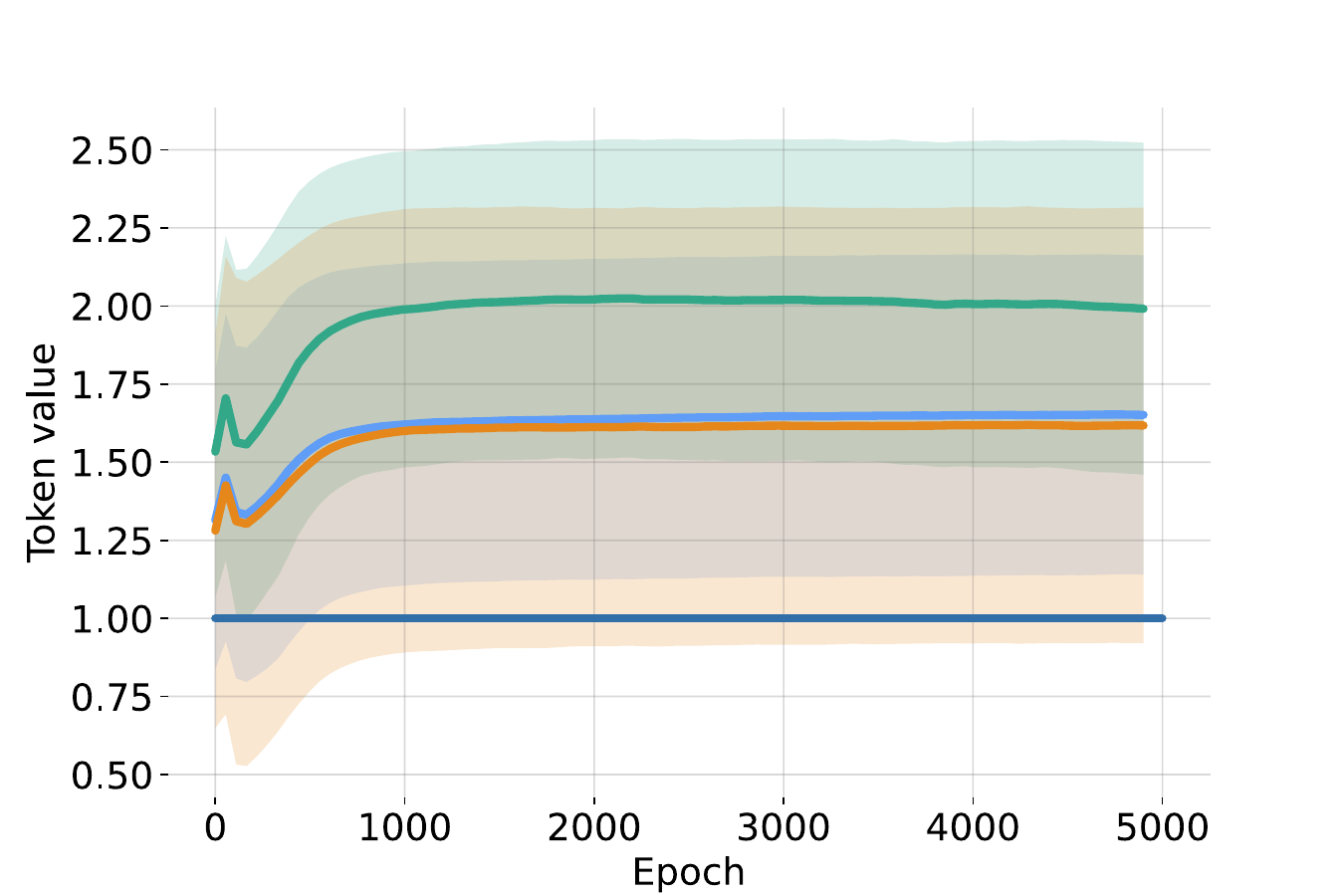}\label{fig:eval:t:coin4}}
  \caption{MEDIATE Evaluation: {\normalfont Comparing the mean \textit{Efficiency} (Fig.~\ref{fig:eval:e:ipd}) and rate of \textit{Own Coins} (Fig.~\ref{fig:eval:e:coin2}, \ref{fig:eval:e:coin4}) of \textit{Naïve Learning} (grey), MATE (blue), AutoMATE (light blue), MEDIATE-I (orange), and MEDIATE-S (green), and the Mean \textit{Token Value} (Fig.~\ref{fig:eval:t:ipd}, \ref{fig:eval:t:coin2}, \ref{fig:eval:t:coin4}) in the \textit{IPD} (Fig.~\ref{fig:eval:e:ipd}, \ref{fig:eval:t:ipd}), \textit{CoinGame-2} (Fig.~\ref{fig:eval:e:coin2}, \ref{fig:eval:t:coin2}), and \textit{CoinGame-4} (Fig.~\ref{fig:eval:e:coin4}, \ref{fig:eval:t:coin4}). The shaded areas mark the 95\% confidence intervals.}}\label{fig:ablation}
\end{figure*}

Fig.~\ref{fig:ablation} shows the evaluation results.
The graphs indicate that either synchronized or isolated MEDIATE updates consistently achieve efficiency and cooperation levels at least equivalent to MATE in all experimental settings, which legitimates their further investigation. 
As expected, naïve learning fails to reach emergent cooperation, again showcasing the compared environments' intricacy. 

In general, MEDIATE enhances the performance of AutoMATE across all settings, except for the two-agent Coin Game scenario, where isolated updates neither improve nor deteriorate cooperation.
The results imply that the combined automatic and decentralized mechanism - introduced by MEDIATE - provides sufficient tokens to replace the original MATE token value $1$.
Furthermore, Figs.~\ref{fig:eval:t:ipd}-\ref{fig:eval:t:coin4} show that all automatically derived tokens converge within the initial 1000 epochs, indicating the purposeful nature of the proposed architecture.
In comparison, the corresponding tokens of AutoMATE and MEDIATE all converge to higher token values than MATE, which, according to our preliminary studies, are more optimal tokens. 
Wider confidence intervals in token convergence are generally associated with reduced efficiency and cooperation, but in the CoinGame-4, AutoMATE tokens converge to equivalent values as those with isolated updates. 
However, although its confidence interval is narrower, AutoMATE's performance is inferior due to the missing token coordination between the agents. 
Comparing the two MEDIATE variants, isolated updates perform better in both CoinGame settings. 

In the negative-valued IPD domain, synchronized updates show advantages. 
Overall, in combination with the token plots, the results show that the update variant converging to a smaller value, i.e., the respectively less optimistic variant, provides superior tokens and thus yields improved efficiency and cooperation. 
Given the absence of a definitive superior option between the two MEDIATE variants, we include both in the benchmark comparisons.

\subsection{Benchmark Comparisons}

\begin{figure*}[t]\centering\includegraphics[width=.6\linewidth]{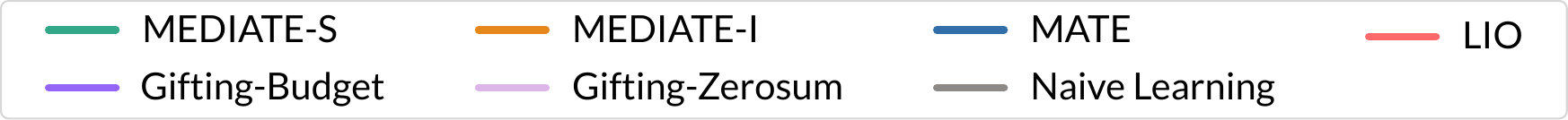}\vspace{-0.3cm}
  \subfloat[\centering RCG-2 | Own Coins]{\includegraphics[width=.33\linewidth]{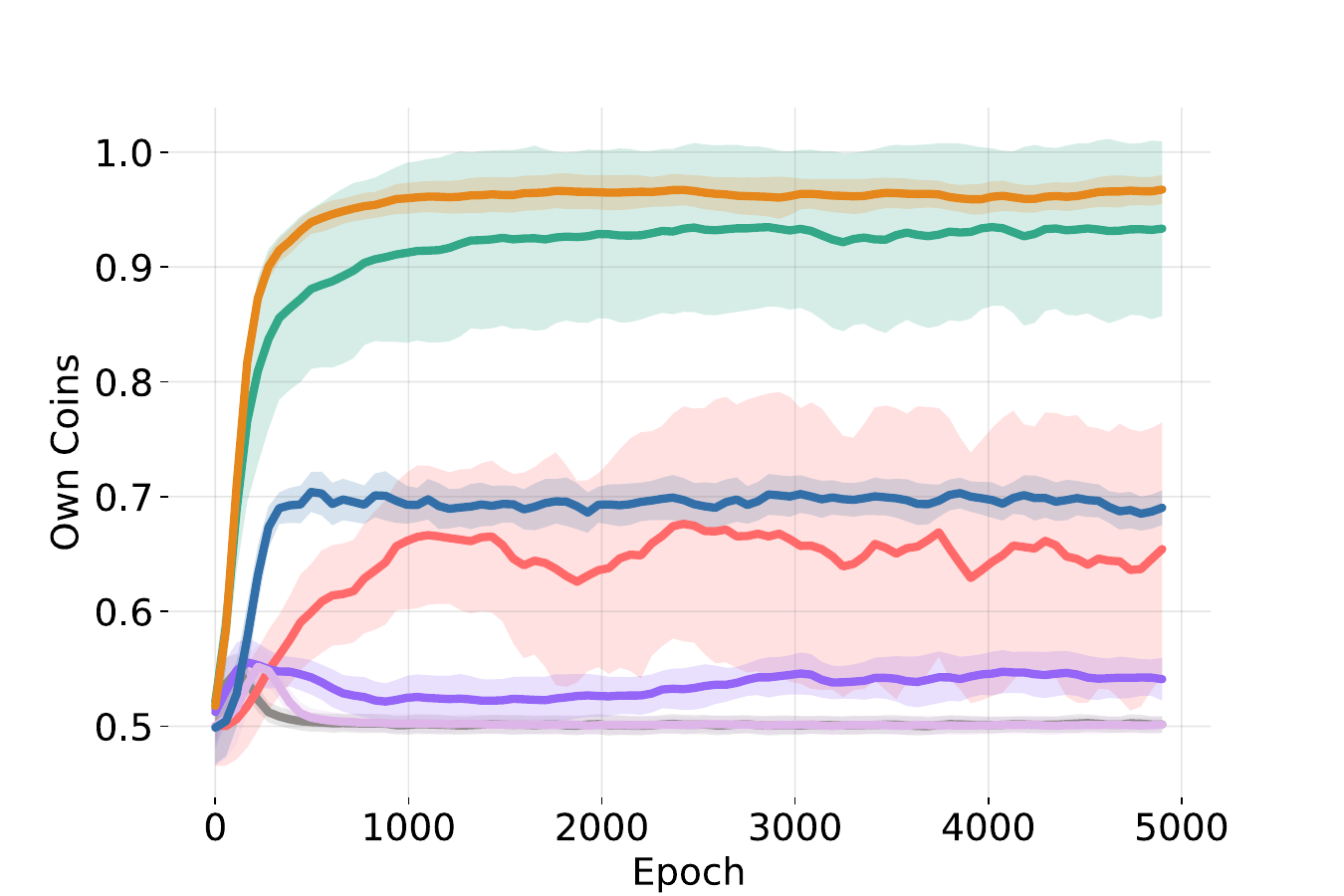} \label{fig:bench:rcoin2}}%
  \subfloat[\centering CG-6 | Own Coins]{\includegraphics[width=.33\linewidth]{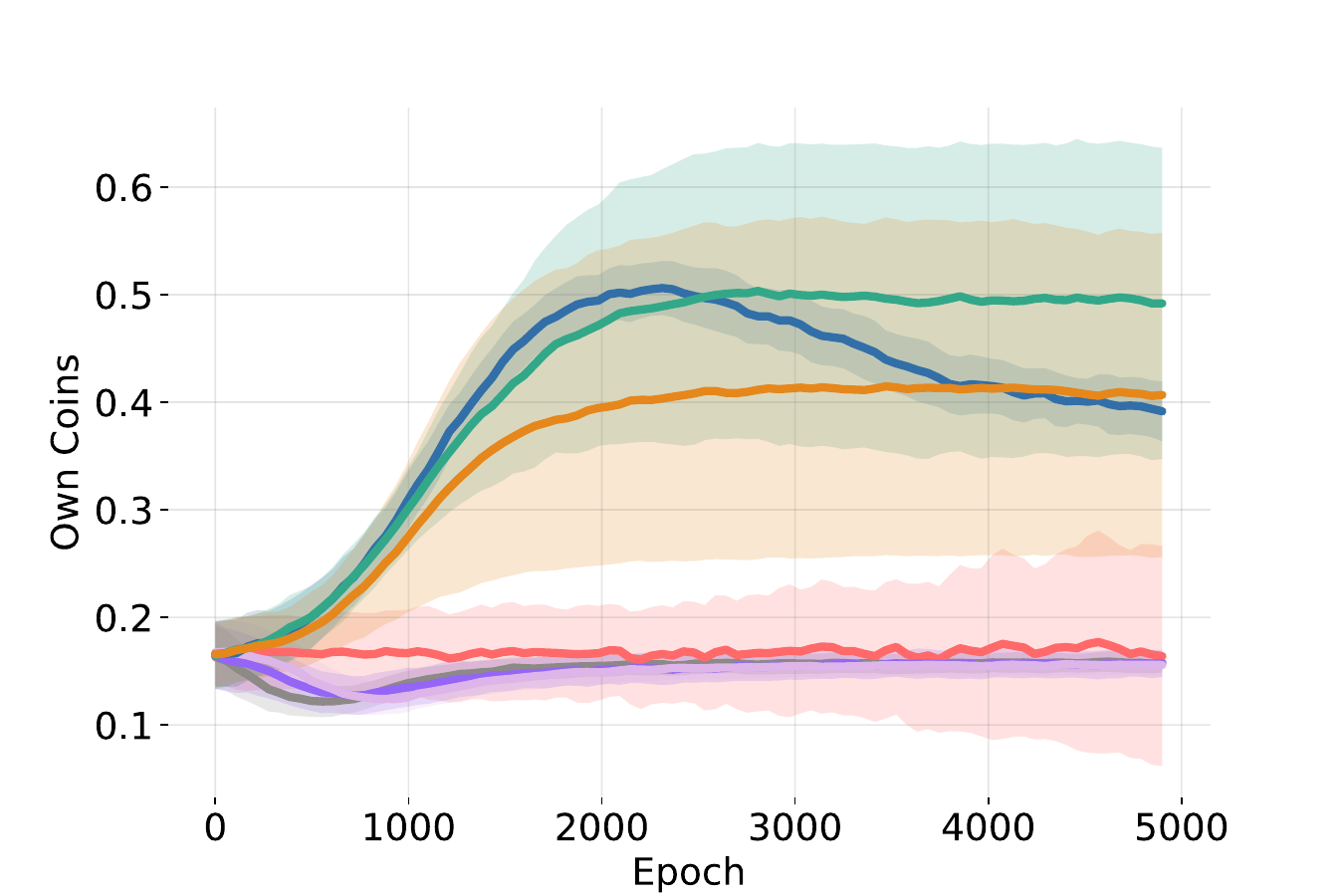}\label{fig:bench:coin6}}%
 \subfloat[\centering Harvest (6 Agents) | Efficiency]{\includegraphics[width=.33\linewidth]{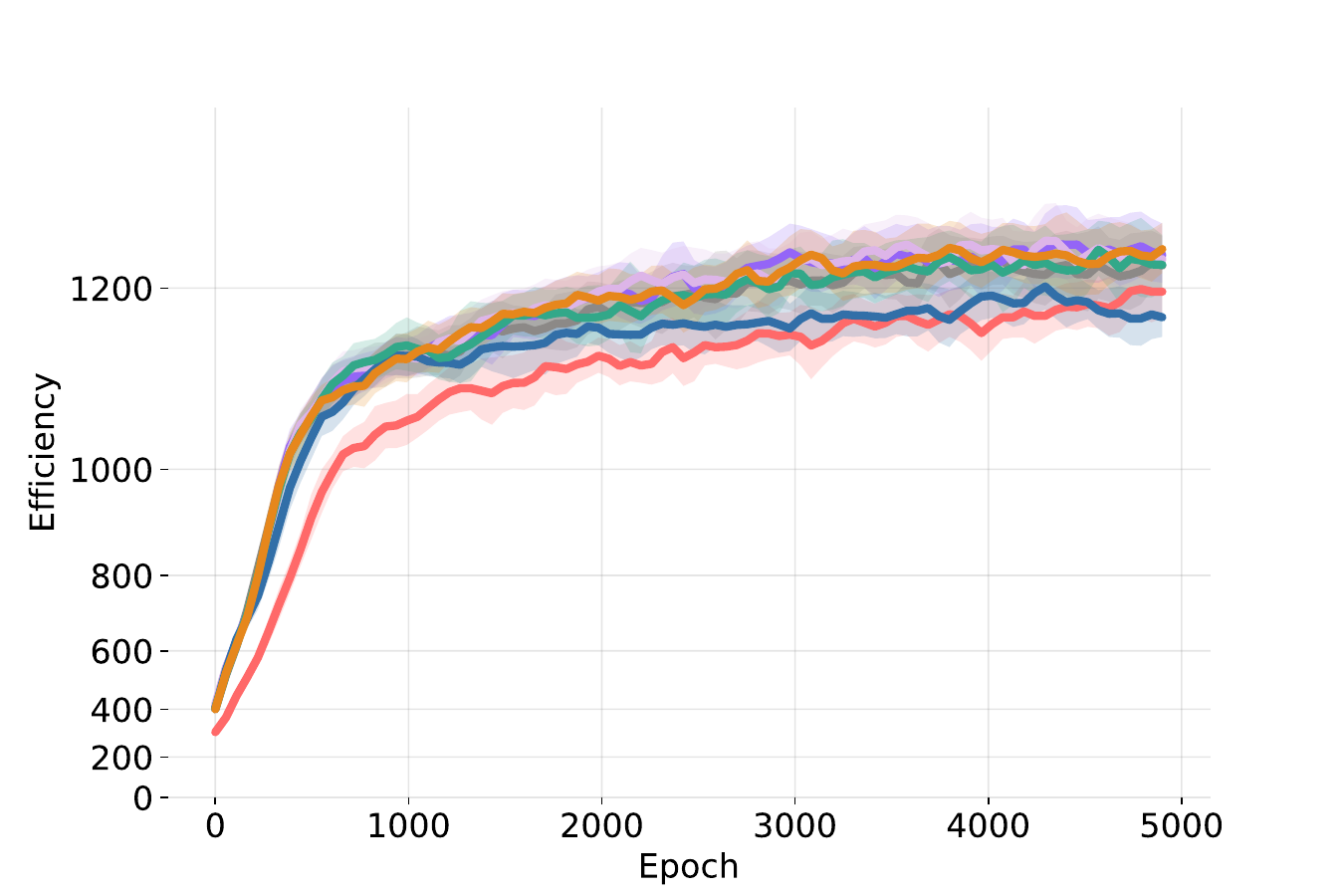}\label{fig:bench:harvest}}
 \caption{Benchmark Comparison: Mean rate of \textit{Own Coins} (Fig.~\ref{fig:bench:rcoin2}, \ref{fig:bench:coin6}) and \textit{Efficiency} (Fig.~\ref{fig:bench:harvest}) of MEDIATE-S (green), MEDIATE-I (orange), MATE (blue), LIO (red), \textit{Budget-Gifting} (purple), \textit{Zerosum-Gifting} (pink) and \textit{Naïve Learning} (grey) in the \textit{Rescaled CoinGame-2} (RCG-2) (Fig.~\ref{fig:bench:rcoin2}), \textit{CoinGame-6} (CG-6) (Fig.~\ref{fig:bench:coin6}), and \textit{Harvest} (Fig.~\ref{fig:bench:harvest}). The shaded areas mark the 95\% confidence intervals.} \label{fig:benchmarks} 
\end{figure*}

Fig.~\ref{fig:benchmarks} shows the benchmark results. 
Table~\ref{tab:benchmark} summarizes the final performance metrics. 
The two-agent Coin Game features down-scaled rewards (RCG-2), requiring agents to learn cooperation under minimal positive and negative environment rewards.
In contrast to the compared approaches, both MEDIATE variants achieve significantly higher rewards and master the task.
Yet, isolated updates exhibit a slight performance advantage over synchronized updates. 
MATE demonstrates moderate cooperation, slightly improving upon LIO. 
In contrast, the gifting methods and naïve learning only show marginal cooperation, although Gifting-Budget performs comparably better.
These results again highlight the superior adaptability of MEDIATE to unconventional, potentially challenging reward scenarios that yield improved applicability to varying tasks. 

In the six-agent Coin Game (CG-6), naïve learning performs worst alongside Gifting-Zerosum and Gifting-Budget. 
While LIO shows a marginal improvement, it still lacks significantly behind MATE and MEDIATE regarding strategic cooperation.
MEDIATE-I performs similar to MATE, which potentially can be attributed to the limited capability of isolated updates to manage negative returns. 
MATE initially demonstrates an optimal learning curve but deteriorates in performance afterward.
In terms of cooperation, MEDIATE with synchronized updates emerges as performing best. 

Harvest demonstrates the ability of MEDIATE to benefit in partially connected topologies. 
Here, MEDIATE ranks among the top-performing approaches and enhances the performance of MATE by providing an appropriate incentivization token. 
It thus demonstrates its efficacy in functioning even within unreliable environments while preserving privacy over the agents' local value information.

\begin{table}[h]\centering 
\caption{Final average of the rate of \textit{Own Coins} in the \textit{Rescaled CoinGame-2} (RCG-2) and \textit{CoinGame-6} (CG-6), and the \textit{Efficiency} in \textit{Harvest} for \textit{synchronized} and \textit{isolated} MEDIATE (MEDIATE-S, MEDIATE-I), AutoMATE, MATE, LIO, \textit{Budget-} and \textit{Zerosum} Gifting (Budget-G, Zerosum-G), and Naïve Learning.}\label{tab:benchmark}
\vskip 0.15in \begin{tabular}{l|c|c|c}
& RCG-2 & CG-6 & Harvest \\
MEDIATE-S  & $0.93 \pm 0.08$                   & $\textbf{0.50} \pm \textbf{0.16}$ & $1212 \pm 20$                  \\ 
MEDIATE-I   & $\textbf{0.97} \pm \textbf{0.02}$ & $0.41 \pm 0.16$                   & $\textbf{1232} \pm \textbf{17}$\\ 
AutoMATE    & $0.86 \pm 0.08$                   & $0.18 \pm 0.09$                   & $1204 \pm 35$                  \\ 
MATE        & $0.69 \pm 0.01$                   & $0.39 \pm 0.03$                   & $\textit{1177} \pm \textit{20}$\\ 
LIO         & $0.69 \pm 0.10$                   & $0.17 \pm 0.11$                   & $1192 \pm 20$                  \\ 
Budget-G    & $0.54 \pm 0.03$                   & $0.16 \pm 0.02$                   & $1232 \pm 23$                  \\ 
Zerosum-G   & $\textit{0.50} \pm \textit{0.01}$ & $\textit{0.16} \pm \textit{0.01}$ & $1230 \pm 20$                  \\ 
Naïve L.    & $\textit{0.50} \pm \textit{0.01}$ & $\textit{0.16} \pm \textit{0.01}$ & $1220 \pm 25$                  \\ 
\end{tabular}\end{table}

Overall, evaluations demonstrated that emergent cooperation between agents fosters optimal social welfare. 
Appropriate reward weights can boost equal cooperation in social dilemmas, but such weights' appropriateness depends on the domain, the number of agents, the reward structure, or other factors. 
Involving a higher number of agents within a domain increases the required precision.
Our experiments show that a token value of 1 - as proposed for MATE - is not universally appropriate in all domains or settings. 
In the down-scaled two-agent Coin Game, token value 1 is inappropriate and in the six-agent Coin Game, it does not achieve optimal cooperation.  
Yet across all domains, MEDIATE exhibits strong adaptability while consistently delivering good performance, even in challenging cooperative tasks such as the six-agent Coin Game, scenarios with complex reward landscapes, or unreliable environments with partially connected neighborhoods, like Harvest.

\section{Conclusion}

In this work, we proposed \textit{mutually endorsed distributed incentive acknowledgment token exchange} (MEDIATE). 
MEDIATE introduces automated PI tokens in decentralized MAS with a consensus architecture and two agent-individual update mechanisms.

Token decentralization allows agents to use different tokens in the exchange. 
Experiments on the impact of different tokens in social dilemmas suggest that equal and appropriate token values foster improved social welfare.
MEDIATE integrates the gradient of the agents' local value function approximation to derive appropriate tokens matching the external rewards.
To achieve consensus on equal tokens, we propose extending the MATE protocol based on additive secret sharing, enabling the identification of the token average through the token exchange while adhering to privacy requirements. 
The consensus protocol is independent of the underlying algorithm for token derivation. 
We furthermore evaluate two token-update variations: A synchronized mechanism based on the reconstructed global token and an isolated mechanism using the previous local token.

Benchmark evaluations showed that MEDIATE achieves high social welfare in all tested domains. 
In all evaluated settings, MEDIATE improves the performance of MATE and even outperforms or matches the best-performing baselines. 
MEDIATE represents a robust and adaptive solution capable of finding appropriate tokens. 
Computationally, MEDIATE is comparable to MATE while overcoming its central limitation of static token values.  
The only addition of deriving consented tokens at each update is a sum of constant values with linear complexity. 
Furthermore, the token extends on the value approximation. 
Thus, compared to LIO, no additional model needs to be learned.

Yet, even though not apparent in the evaluated social dilemma environments, this dependence on a robust value estimate also depicts a central limitation of MEDIATE. 
Furthermore, the evaluated update mechanisms showed potentially unstable and prone to outliers. 
Thus, future work should focus on producing more accurate tokens, especially for an increased number of agents, making the overall algorithm more reliable in precision-requiring domains like the Rescaled CoinGame. 
Also, while shown robust to scaled reward landscapes, increasing numbers of agents, and long-term cooperation scenarios like Harvest, MEDIATE should be tested for unreliable connections or defective scenarios. 

Overall, MEDIATE provides a lightweight and robust framework to assess communication consensus mechanisms with automated peer incentives for creating emergent cooperation in various scenarios of social dilemmas.



\section*{Impact Statement}
MEDIATE represents a protocol addition to derive decentralized token values for current peer incentivization approaches using a privacy-preserving consensus mechanism. 
Using those improved token values overall increases systematic cooperation between self-interested agents. 
Even though, in theory, the system is designed to be robust w.r.t. defectors, additional evaluations with adversaries would be helpful. 





\begin{thebibliography}{47}
\providecommand{\natexlab}[1]{#1}
\providecommand{\url}[1]{\texttt{#1}}
\expandafter\ifx\csname urlstyle\endcsname\relax
  \providecommand{\doi}[1]{doi: #1}\else
  \providecommand{\doi}{doi: \begingroup \urlstyle{rm}\Url}\fi

\bibitem[Agarap(2018)]{agarap2018deep}
Agarap, A.~F.
\newblock Deep learning using rectified linear units (relu).
\newblock \emph{arXiv preprint arXiv:1803.08375}, 2018.

\bibitem[Amirkhani \& Barshooi(2022)Amirkhani and Barshooi]{amirkhani2022consensus}
Amirkhani, A. and Barshooi, A.~H.
\newblock Consensus in multi-agent systems: a review.
\newblock \emph{Artificial Intelligence Review}, 55\penalty0 (5):\penalty0 3897--3935, 2022.

\bibitem[Axelrod(1980)]{axelrod1980effective}
Axelrod, R.
\newblock Effective choice in the prisoner's dilemma.
\newblock \emph{Journal of conflict resolution}, 24\penalty0 (1):\penalty0 3--25, 1980.

\bibitem[Baumann et~al.(2020)Baumann, Graepel, and Shawe-Taylor]{baumann2020adaptive}
Baumann, T., Graepel, T., and Shawe-Taylor, J.
\newblock Adaptive mechanism design: Learning to promote cooperation.
\newblock In \emph{2020 International Joint Conference on Neural Networks (IJCNN)}, pp.\  1--7. IEEE, 2020.

\bibitem[Bu{\c{s}}oniu et~al.(2010)Bu{\c{s}}oniu, Babu{\v{s}}ka, and De~Schutter]{bucsoniu2010multi}
Bu{\c{s}}oniu, L., Babu{\v{s}}ka, R., and De~Schutter, B.
\newblock Multi-agent reinforcement learning: An overview.
\newblock \emph{Innovations in multi-agent systems and applications-1}, pp.\  183--221, 2010.

\bibitem[Cao et~al.(2019)Cao, Li, Zhang, Zhang, Mumtaz, Zhou, and Peng]{cao2019internet}
Cao, B., Li, Y., Zhang, L., Zhang, L., Mumtaz, S., Zhou, Z., and Peng, M.
\newblock When internet of things meets blockchain: Challenges in distributed consensus.
\newblock \emph{IEEE Network}, 33\penalty0 (6):\penalty0 133--139, 2019.

\bibitem[Conradt \& Roper(2005)Conradt and Roper]{conradt2005consensus}
Conradt, L. and Roper, T.~J.
\newblock Consensus decision making in animals.
\newblock \emph{Trends in ecology \& evolution}, 20\penalty0 (8):\penalty0 449--456, 2005.

\bibitem[Crainic \& Toulouse(2007)Crainic and Toulouse]{crainic2007explicit}
Crainic, T.~G. and Toulouse, M.
\newblock Explicit and emergent cooperation schemes for search algorithms.
\newblock In \emph{International Conference on Learning and Intelligent Optimization}, pp.\  95--109. Springer, 2007.

\bibitem[Dawes(1980)]{dawes1980social}
Dawes, R.~M.
\newblock Social dilemmas.
\newblock \emph{Annual review of psychology}, 31\penalty0 (1):\penalty0 169--193, 1980.

\bibitem[Eccles et~al.(2019)Eccles, Hughes, Kram{\'a}r, Wheelwright, and Leibo]{eccles2019learning}
Eccles, T., Hughes, E., Kram{\'a}r, J., Wheelwright, S., and Leibo, J.~Z.
\newblock Learning reciprocity in complex sequential social dilemmas.
\newblock \emph{arXiv preprint arXiv:1903.08082}, 2019.

\bibitem[Fayad \& Ibrahim(2021)Fayad and Ibrahim]{fayad2021influence}
Fayad, A. and Ibrahim, M.
\newblock Influence-based reinforcement learning for intrinsically-motivated agents.
\newblock \emph{arXiv preprint arXiv:2108.12581}, 2021.

\bibitem[Figura et~al.(2021)Figura, Kosaraju, and Gupta]{figura2021adversarial}
Figura, M., Kosaraju, K.~C., and Gupta, V.
\newblock Adversarial attacks in consensus-based multi-agent reinforcement learning.
\newblock In \emph{2021 American Control Conference (ACC)}, pp.\  3050--3055. IEEE, 2021.

\bibitem[Foerster et~al.(2018)Foerster, Chen, Al-Shedivat, Whiteson, Abbeel, and Mordatch]{Foerster2017Learning}
Foerster, J., Chen, R.~Y., Al-Shedivat, M., Whiteson, S., Abbeel, P., and Mordatch, I.
\newblock Learning with opponent-learning awareness.
\newblock In \emph{Proceedings of the 17th International Conference on Autonomous Agents and MultiAgent Systems}, pp.\  122–130, 2018.

\bibitem[Han et~al.(2013)Han, Lu, and Chen]{han2013cluster}
Han, Y., Lu, W., and Chen, T.
\newblock Cluster consensus in discrete-time networks of multiagents with inter-cluster nonidentical inputs.
\newblock \emph{IEEE Transactions on Neural Networks and Learning Systems}, 24\penalty0 (4):\penalty0 566--578, 2013.

\bibitem[Hostallero et~al.(2020)Hostallero, Kim, Moon, Son, Kang, and Yi]{hostallero2020inducing}
Hostallero, D.~E., Kim, D., Moon, S., Son, K., Kang, W.~J., and Yi, Y.
\newblock Inducing cooperation through reward reshaping based on peer evaluations in deep multi-agent reinforcement learning.
\newblock In \emph{Proceedings of the 19th International Conference on Autonomous Agents and MultiAgent Systems}, pp.\  520--528, 2020.

\bibitem[Jaques et~al.(2019)Jaques, Lazaridou, Hughes, Gulcehre, Ortega, Strouse, Leibo, and De~Freitas]{jaques2019social}
Jaques, N., Lazaridou, A., Hughes, E., Gulcehre, C., Ortega, P., Strouse, D., Leibo, J.~Z., and De~Freitas, N.
\newblock Social influence as intrinsic motivation for multi-agent deep reinforcement learning.
\newblock In \emph{International conference on machine learning}, pp.\  3040--3049. PMLR, 2019.

\bibitem[Kim et~al.(2020)Kim, Lee, Son, Bae, and Do~Chung]{kim2020multi}
Kim, Y.~G., Lee, S., Son, J., Bae, H., and Do~Chung, B.
\newblock Multi-agent system and reinforcement learning approach for distributed intelligence in a flexible smart manufacturing system.
\newblock \emph{Journal of Manufacturing Systems}, 57:\penalty0 440--450, 2020.

\bibitem[Kuhnle et~al.(2023)Kuhnle, Richley, and Perez-Lavin]{kuhnle2023learning}
Kuhnle, A., Richley, J., and Perez-Lavin, D.
\newblock Learning strategic value and cooperation in multi-player stochastic games through side payments.
\newblock \emph{arXiv preprint arXiv:2303.05307}, 2023.

\bibitem[Lashkari \& Musilek(2021)Lashkari and Musilek]{lashkari2021comprehensive}
Lashkari, B. and Musilek, P.
\newblock A comprehensive review of blockchain consensus mechanisms.
\newblock \emph{IEEE Access}, 9:\penalty0 43620--43652, 2021.

\bibitem[Laurent et~al.(2011)Laurent, Matignon, Fort-Piat, et~al.]{laurent2011world}
Laurent, G.~J., Matignon, L., Fort-Piat, L., et~al.
\newblock The world of independent learners is not markovian.
\newblock \emph{International Journal of Knowledge-based and Intelligent Engineering Systems}, 15\penalty0 (1):\penalty0 55--64, 2011.

\bibitem[Leibo et~al.(2017)Leibo, Zambaldi, Lanctot, Marecki, and Graepel]{leibo2017multi}
Leibo, J.~Z., Zambaldi, V., Lanctot, M., Marecki, J., and Graepel, T.
\newblock Multi-agent reinforcement learning in sequential social dilemmas.
\newblock In \emph{Proceedings of the 16th Conference on Autonomous Agents and MultiAgent Systems}, pp.\  464--473, 2017.

\bibitem[Lerer \& Peysakhovich(2017)Lerer and Peysakhovich]{lerer2017maintaining}
Lerer, A. and Peysakhovich, A.
\newblock Maintaining cooperation in complex social dilemmas using deep reinforcement learning.
\newblock \emph{arXiv preprint arXiv:1707.01068}, 2017.

\bibitem[Letcher et~al.(2019)Letcher, Foerster, Balduzzi, Rocktäschel, and Whiteson]{letcher2018stable}
Letcher, A., Foerster, J., Balduzzi, D., Rocktäschel, T., and Whiteson, S.
\newblock Stable opponent shaping in differentiable games.
\newblock In \emph{International Conference on Learning Representations}, 2019.

\bibitem[Li et~al.(2019)Li, Cascudo, and Christensen]{li2019privacy}
Li, Q., Cascudo, I., and Christensen, M.~G.
\newblock Privacy-preserving distributed average consensus based on additive secret sharing.
\newblock In \emph{2019 27th European Signal Processing Conference (EUSIPCO)}, pp.\  1--5. IEEE, 2019.

\bibitem[Li \& Tan(2019)Li and Tan]{Li2019ASO}
Li, Y. and Tan, C.
\newblock A survey of the consensus for multi-agent systems.
\newblock \emph{Systems Science \& Control Engineering}, 7:\penalty0 468 -- 482, 2019.

\bibitem[Littman(2001)]{littman2001value}
Littman, M.~L.
\newblock Value-function reinforcement learning in markov games.
\newblock \emph{Cognitive systems research}, 2\penalty0 (1):\penalty0 55--66, 2001.

\bibitem[Lupu \& Precup(2020)Lupu and Precup]{lupu2020gifting}
Lupu, A. and Precup, D.
\newblock Gifting in multi-agent reinforcement learning.
\newblock In \emph{Proceedings of the 19th International Conference on autonomous agents and multiagent systems}, pp.\  789--797, 2020.

\bibitem[Merhej \& Chetouani(2021)Merhej and Chetouani]{merhej2021lief}
Merhej, R. and Chetouani, M.
\newblock Lief: Learning to influence through evaluative feedback.
\newblock In \emph{Adaptive and Learning Agents Workshop (AAMAS 2021)}, 2021.

\bibitem[Monrat et~al.(2019)Monrat, Schel{\'e}n, and Andersson]{monrat2019survey}
Monrat, A.~A., Schel{\'e}n, O., and Andersson, K.
\newblock A survey of blockchain from the perspectives of applications, challenges, and opportunities.
\newblock \emph{IEEE Access}, 7:\penalty0 117134--117151, 2019.

\bibitem[No{\"e}(2006)]{noe2006cooperation}
No{\"e}, R.
\newblock Cooperation experiments: coordination through communication versus acting apart together.
\newblock \emph{Animal behaviour}, 71\penalty0 (1):\penalty0 1--18, 2006.

\bibitem[Olfati-Saber \& Shamma(2005)Olfati-Saber and Shamma]{olfati2005consensus}
Olfati-Saber, R. and Shamma, J.~S.
\newblock Consensus filters for sensor networks and distributed sensor fusion.
\newblock In \emph{Proceedings of the 44th IEEE Conference on Decision and Control}, pp.\  6698--6703. IEEE, 2005.

\bibitem[Omitaomu \& Niu(2021)Omitaomu and Niu]{omitaomu2021artificial}
Omitaomu, O.~A. and Niu, H.
\newblock Artificial intelligence techniques in smart grid: A survey.
\newblock \emph{Smart Cities}, 4\penalty0 (2):\penalty0 548--568, 2021.

\bibitem[Perolat et~al.(2017)Perolat, Leibo, Zambaldi, Beattie, Tuyls, and Graepel]{perolat2017multiagent}
Perolat, J., Leibo, J.~Z., Zambaldi, V., Beattie, C., Tuyls, K., and Graepel, T.
\newblock A multi-agent reinforcement learning model of common-pool resource appropriation, 2017.

\bibitem[Phan et~al.(2022)Phan, Sommer, Altmann, Ritz, Belzner, and Linnhoff-Popien]{phan2022emergent}
Phan, T., Sommer, F., Altmann, P., Ritz, F., Belzner, L., and Linnhoff-Popien, C.
\newblock Emergent cooperation from mutual acknowledgment exchange.
\newblock In \emph{Proceedings of the 21st International Conference on Autonomous Agents and Multiagent Systems}, pp.\  1047--1055, 2022.

\bibitem[Qureshi \& Abdullah(2013)Qureshi and Abdullah]{qureshi2013survey}
Qureshi, K.~N. and Abdullah, A.~H.
\newblock A survey on intelligent transportation systems.
\newblock \emph{Middle-East Journal of Scientific Research}, 15\penalty0 (5):\penalty0 629--642, 2013.

\bibitem[Rashid et~al.(2020)Rashid, Samvelyan, De~Witt, Farquhar, Foerster, and Whiteson]{rashid2020monotonic}
Rashid, T., Samvelyan, M., De~Witt, C.~S., Farquhar, G., Foerster, J., and Whiteson, S.
\newblock Monotonic value function factorisation for deep multi-agent reinforcement learning.
\newblock \emph{The Journal of Machine Learning Research}, 21\penalty0 (1):\penalty0 7234--7284, 2020.

\bibitem[Russell(2010)]{russell2010artificial}
Russell, S.~J.
\newblock \emph{Artificial intelligence a modern approach}.
\newblock Pearson Education, Inc., 2010.

\bibitem[Salimitari \& Chatterjee(2018)Salimitari and Chatterjee]{salimitari2018survey}
Salimitari, M. and Chatterjee, M.
\newblock A survey on consensus protocols in blockchain for iot networks.
\newblock \emph{arXiv preprint arXiv:1809.05613}, 2018.

\bibitem[Sandholm \& Crites(1996)Sandholm and Crites]{sandholm1996multiagent}
Sandholm, T.~W. and Crites, R.~H.
\newblock Multiagent reinforcement learning in the iterated prisoner's dilemma.
\newblock \emph{Biosystems}, 37\penalty0 (1-2):\penalty0 147--166, 1996.

\bibitem[Schenato \& Gamba(2007)Schenato and Gamba]{schenato2007distributed}
Schenato, L. and Gamba, G.
\newblock A distributed consensus protocol for clock synchronization in wireless sensor network.
\newblock In \emph{2007 46th ieee conference on decision and control}, pp.\  2289--2294. IEEE, 2007.

\bibitem[Son et~al.(2019)Son, Kim, Kang, Hostallero, and Yi]{son2019qtran}
Son, K., Kim, D., Kang, W.~J., Hostallero, D.~E., and Yi, Y.
\newblock Qtran: Learning to factorize with transformation for cooperative multi-agent reinforcement learning.
\newblock In \emph{International conference on machine learning}, pp.\  5887--5896. PMLR, 2019.

\bibitem[Sunehag et~al.(2018)Sunehag, Lever, Gruslys, Czarnecki, Zambaldi, Jaderberg, Lanctot, Sonnerat, Leibo, Tuyls, and Graepel]{Sunehag2018}
Sunehag, P., Lever, G., Gruslys, A., Czarnecki, W.~M., Zambaldi, V., Jaderberg, M., Lanctot, M., Sonnerat, N., Leibo, J.~Z., Tuyls, K., and Graepel, T.
\newblock Value-decomposition networks for cooperative multi-agent learning based on team reward.
\newblock In \emph{Proceedings of the 17th International Conference on Autonomous Agents and MultiAgent Systems}, pp.\  2085–2087, 2018.

\bibitem[Tawalbeh et~al.(2020)Tawalbeh, Muheidat, Tawalbeh, and Quwaider]{tawalbeh2020iot}
Tawalbeh, L., Muheidat, F., Tawalbeh, M., and Quwaider, M.
\newblock Iot privacy and security: Challenges and solutions.
\newblock \emph{Applied Sciences}, 10\penalty0 (12):\penalty0 4102, 2020.

\bibitem[Wang et~al.(2019)Wang, Hughes, Fernando, Czarnecki, Du\'{e}\~{n}ez Guzm\'{a}n, and Leibo]{Wang2018Evolving}
Wang, J.~X., Hughes, E., Fernando, C., Czarnecki, W.~M., Du\'{e}\~{n}ez Guzm\'{a}n, E.~A., and Leibo, J.~Z.
\newblock Evolving intrinsic motivations for altruistic behavior.
\newblock In \emph{Proceedings of the 18th International Conference on Autonomous Agents and MultiAgent Systems}, pp.\  683–692, 2019.

\bibitem[Yang et~al.(2020)Yang, Li, Farajtabar, Sunehag, Hughes, and Zha]{yang2020learning}
Yang, J., Li, A., Farajtabar, M., Sunehag, P., Hughes, E., and Zha, H.
\newblock Learning to incentivize other learning agents.
\newblock \emph{Advances in Neural Information Processing Systems}, 33:\penalty0 15208--15219, 2020.

\bibitem[Yi et~al.(2021)Yi, Li, Wang, and Lu]{yi2021learning}
Yi, Y., Li, G., Wang, Y., and Lu, Z.
\newblock Learning to share in multi-agent reinforcement learning.
\newblock \emph{arXiv preprint arXiv:2112.08702}, 2021.

\bibitem[Yu et~al.(2009)Yu, Chen, Wang, and Yang]{yu2009distributed}
Yu, W., Chen, G., Wang, Z., and Yang, W.
\newblock Distributed consensus filtering in sensor networks.
\newblock \emph{IEEE Transactions on Systems, Man, and Cybernetics, Part B (Cybernetics)}, 39\penalty0 (6):\penalty0 1568--1577, 2009.

\end{thebibliography}

\newpage
\appendix
\onecolumn



\section{Environments}\label{sec:envs}

\paragraph{Iterated Prisoner's Dilemma}\label{subsubsec:IPD}

\begin{wraptable}{r}{8cm}
  \vspace{-0.5cm}
  \centering 
  \caption{Prisoner's Dilemma reward allocation. Each cell contains the respective payoffs for each of the two players based on their choice of cooperation or defection.}
  \vspace{0.5cm}
  \label{tab:ipd}
  \begin{tabular}{|p{2cm}|p{2cm}|p{2cm}|}
  \cline{2-3}
    \multicolumn{1}{c|}{} & \textbf{Cooperate} & \textbf{Defect} \\
    \cline{1-3}
    \textbf{Cooperate} & (-1,-1) & (-3,0) \\
    \cline{1-3}
    \textbf{Defect} & (0,-3) & (-2,-2) \\
    \cline{1-3}
  \end{tabular}
\end{wraptable}

The \textit{Iterated Prisoner's Dilemma} (IPD) is the repeated game of the Prisoner's Dilemma, depicted in Table~\ref{tab:ipd}. At each time step, the two players must choose between cooperation and defection to maximize their payoff \cite{axelrod1980effective, hostallero2020inducing}. Mutual defection constitutes a Nash equilibrium. 
If both agents defect, no agent is incentivized to change its strategy to cooperation in the next step if the other agent remains a defector. 
If both agents switched their strategy to cooperate, both would receive a lower penalty.

\paragraph{Coin Game}

\begin{figure}[b]
  \centering
   \subfloat[\textit{CoinGame-$2$}]{\includegraphics[width=.15\linewidth]{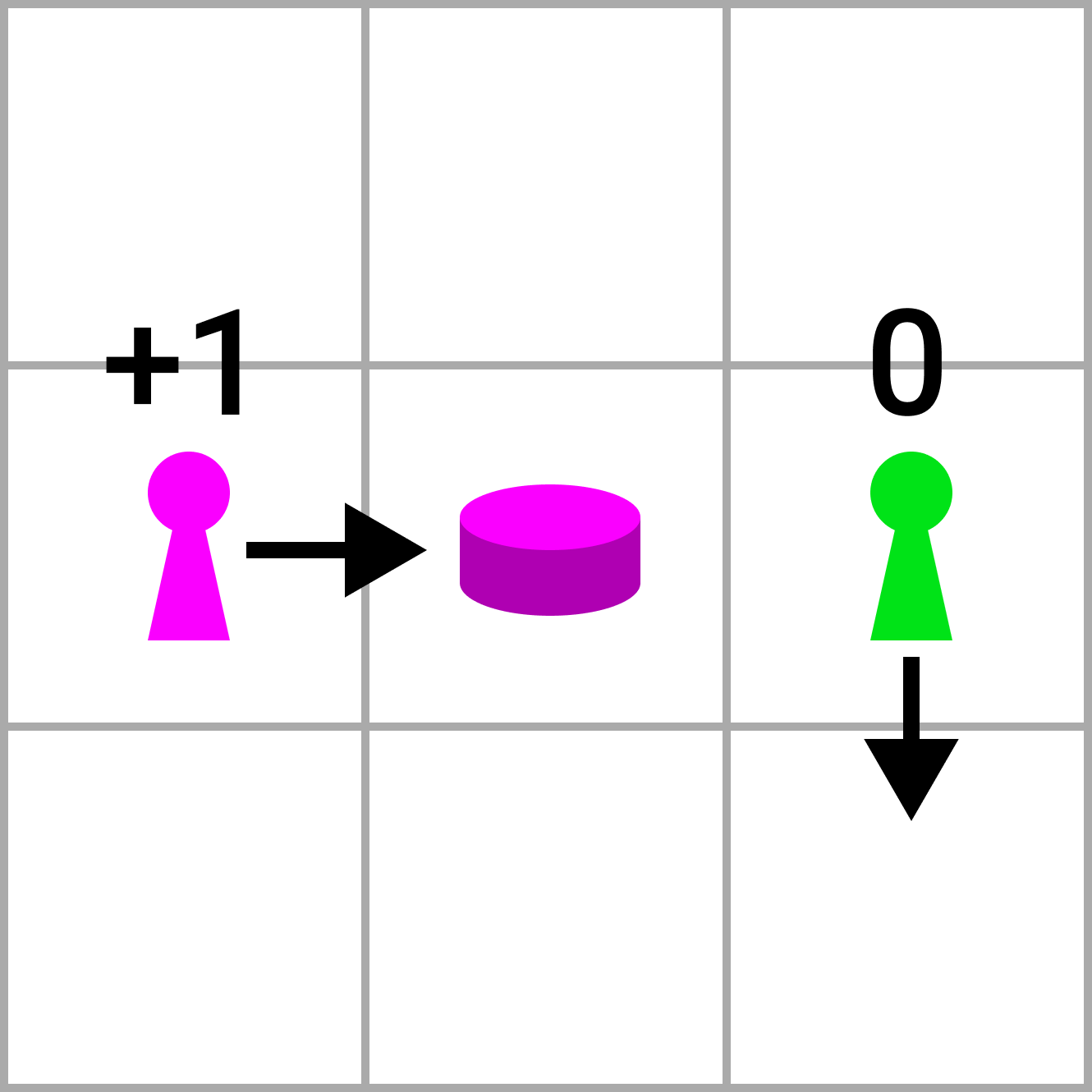}\label{fig:env:coin2}}\quad
   \subfloat[\textit{CoinGame-$4$}]{\includegraphics[width=.15\linewidth]{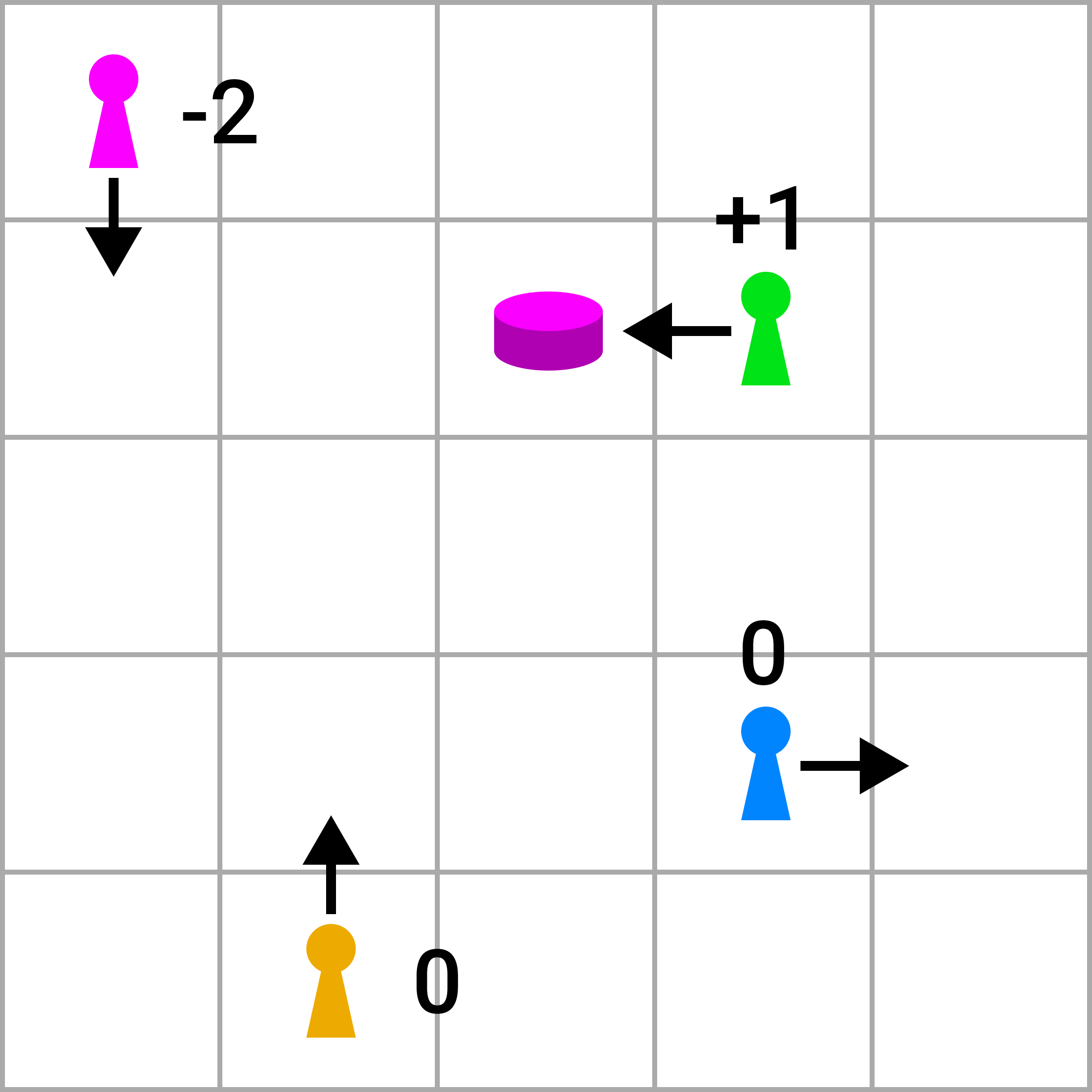}\label{fig:env:coin4}}\quad
   \subfloat[\textit{CoinGame-$6$}]{\includegraphics[width=.15\linewidth]{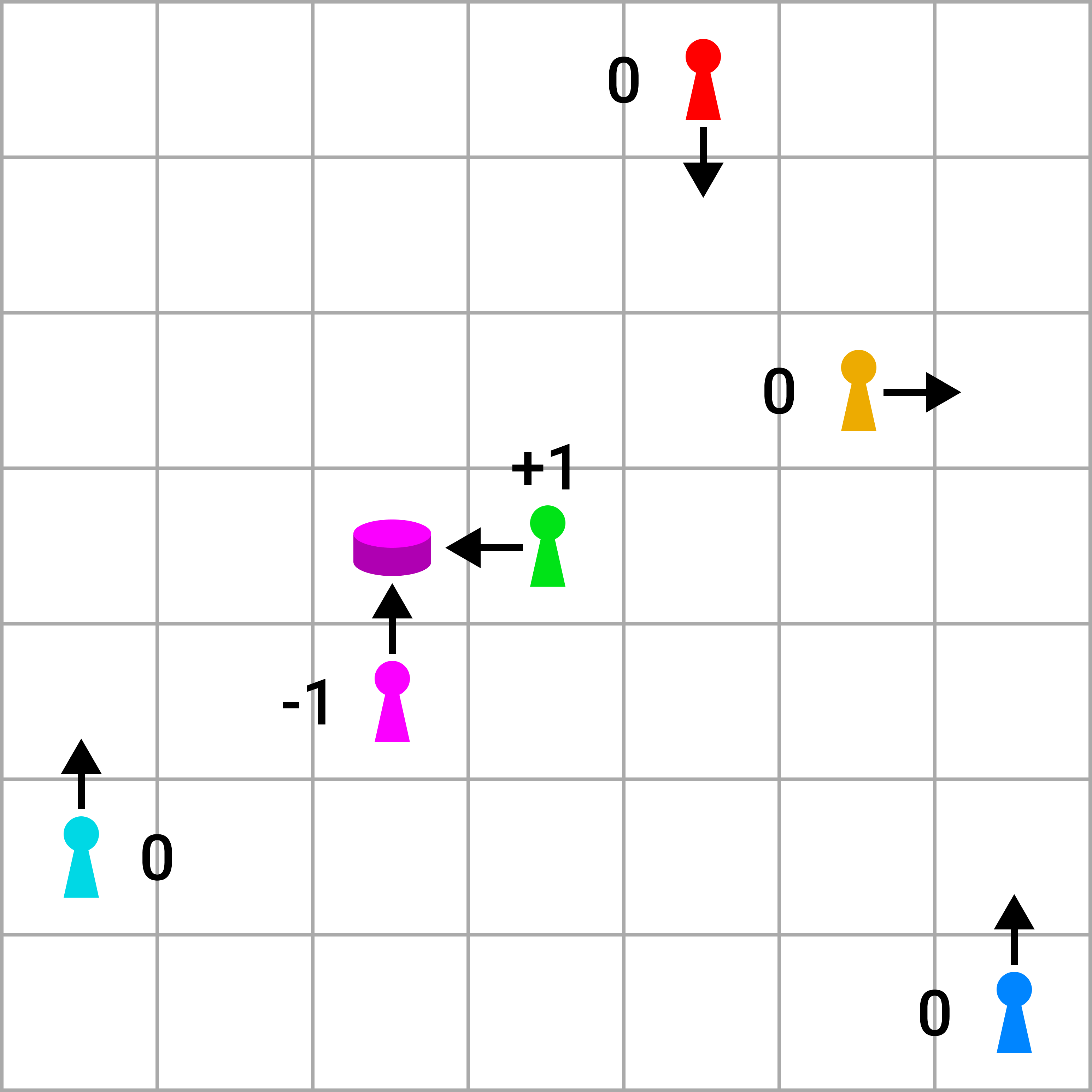}\label{fig:env:coin6}}\quad   \subfloat[Harvest (6 agents)]{\includegraphics[width=.42\linewidth]{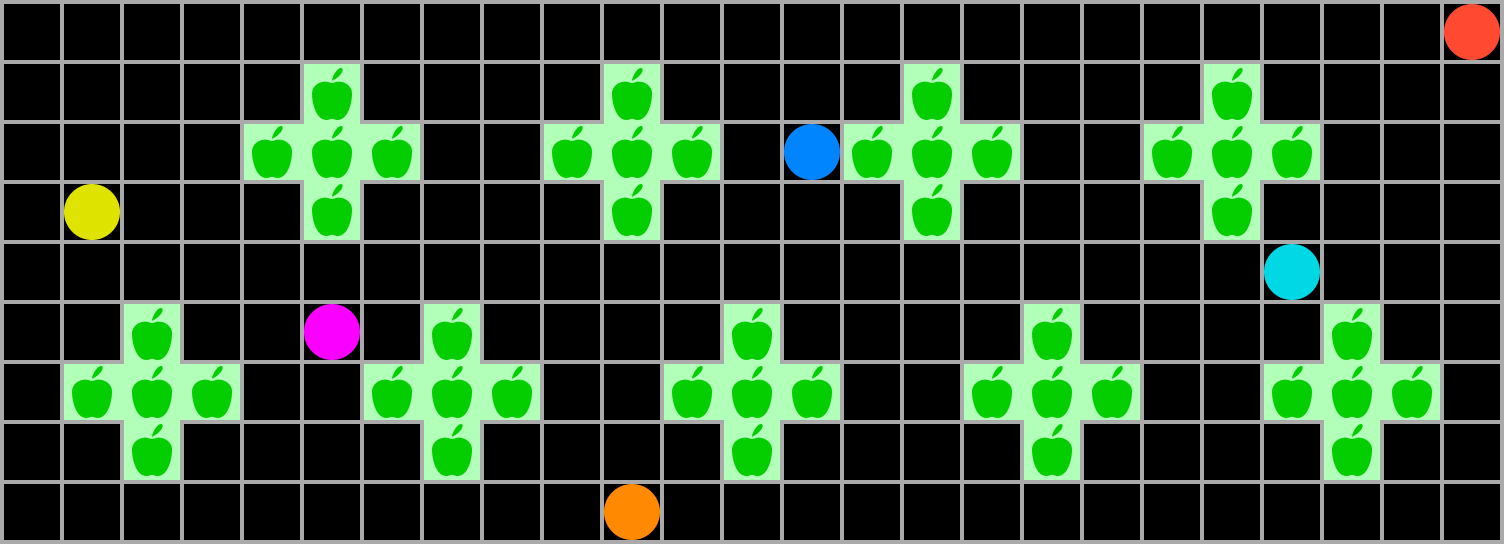}\label{fig:env:harvest-map}}%
\caption{Evaluation Environments} \label{fig:envs}
\end{figure}

Coins or Coin Game is an SSD conceptualized by \citet{lerer2017maintaining}. 
The \textit{CoinGame-$N$} comprises $N \in \{ 2, 4, 6\}$ agents on a $3x3$, $5x5$, and $7x7$ grid respectively (cf. Figs.~\ref{fig:env:coin2}-\ref{fig:env:coin4}).
A distinct color identifies each agent. 
Initially, all $N$ agents and one random-colored coin spawn at random positions. 
The color of the coin matches one of the agents.
An agent can distinguish whether the coin matches its own color or not. 
The action space of each agent comprises four directions of movement $\mathcal{A}\in\{ left, right, up, down\}$.
A coin is collected when an agent moves to its position. 
The environment discards actions violating its bounds. 
If an agent collects any coin, it receives a reward of $+1$. 
If the color matches a different agent, that agent is penalized with $-2$. 
If multiple agents collect a coin simultaneously, the matching agent receives a penalty of $-1$. 
Once a coin is collected, a new coin spawns.
To evaluate varying reward scales, we added the \textit{Rescaled Coin Game-$2$} variation with downsized rewards (i.e., scaled by $0.1$), such that the positive reward becomes $+0.1$ and the penalty weighs $-0.2$. 
The ratio between reward and penalty remains unchanged. 
Self-interested agents will collect all coins regardless of color since this strategy imposes only positive rewards on themselves. 
The Nash equilibrium is reached if all agents follow this strategy since refraining from collecting other agents' coins only reduces an agent's own rewards without mitigating the penalties incurred from the actions of other agents. 
However, if all agents collect their own coins, each agent profits from the reduced penalties, and social welfare can be maximized.
To measure the level of strategic cooperation in this domain, we evaluate the rate of \textit{own coins} w.r.t. to the total of collected coins.

\paragraph{Harvest}\label{subsubsec:Harvest}

The Commons game is conceptualized by \citet{perolat2017multiagent} and adapted by \citet{phan2022emergent}, where it is named Harvest. 
In Harvest, agents move on a 25x9-sized grid to collect apples. 
The Harvest grid, including the fixed positioning of the apples, is displayed in Fig.~\ref{fig:env:harvest-map}. 
Apples have a regrowth rate, which depends on the number of existing apples in the local area. 
More apples in the area cause a higher regrowth rate of collected apples. 
If no apples remain in the area, no apples regrow. 
Self-interested agents maximize their own apple harvest, but in a MAS, agents have to refrain from simultaneous apple collection to avoid the ultimate depletion of resources (the \textit{tragedy of the commons}). 
This requirement is the Nash equilibrium of Harvest, as a single agent can not improve its rewards by refraining from apple collection when other agents will continue to diminish the resources. 
Only if all agents cooperate they can maximize their long-term rewards. 
Agents can tag other agents to remove them from the game for 25 time steps. \cite{perolat2017multiagent}. 
In addition to a positive reward of $+1$ for an apple harvest, each time step poses a time penalty of $-0.1$. 
Furthermore, agents only have access to a partial observation surrounding their position. 
Agents can only communicate with agents in their neighborhood in an area of $7x7$ tiles. 
In addition to moving in four directions (as for the coin game), the action space comprises four actions to tag all neighbor agents in the four directions. 
Moving toward a boundary results in no movement. 
Only one agent can harvest an apple or tag another agent at a time. 
The order of actions at each time step is random. 

\section{Implementation Details}


We use the following hyperparameters:

\begin{table}[h]
    \centering
    \caption{Environment Parameters}
    \vskip 0.15in
    \begin{tabular}{l|c|c|c|c|c}\toprule
        & IPD & CoinGame2 & CoinGame4 & CoinGame6 & Harvest  \\ \midrule
       Num actions $|\mathcal{A}|$& 2 & 4 & 4 & 4 & 9 \\
       Observation dimension $|\mathcal{Z}|$ & 2 & 36 & 100 & 196 & 196 \\       
       Discount factor $\gamma$ & $0.95$ & $0.95$ & $0.95$ & $0.95$ & $0.99$ \\
       Time Limit & 150 & 150 & 150 & 150 & 250 \\
       LIO max incentive & 3 & 2 & 2 & 2 & 0.25 \\
    \bottomrule\end{tabular}
\end{table}

\begin{table}[h]
    \centering
    \caption{Training Parameters}
    \vskip 0.15in
    \begin{tabular}{c|c}\toprule
       Name  & Value  \\ \midrule
       Epochs  & $5000$ \\
       Episodes per epoch & $10$ \\
       Network architecture & Shared fully connected prepossessing net: $|\mathcal{Z}|\mapsto64$ \\
       Activation function & Exponential Linear Unit (ELU) \\
       Hidden size & $\{64,64\}$ \\
       Actor head & $64\mapsto |\mathcal{A}|$ (Softmax) \\
       Critic head & $64\mapsto1$ \\
       Optimizer & Adam \\
       Learning rate &  $0.001$ \\
       Clipping norm & $1$ \\
       History length & $1$ \\
    \bottomrule\end{tabular}
\end{table}




\end{document}